\documentclass[10pt,journal]{IEEEtran}
\usepackage{amsmath,amssymb,epsfig,color}
\allowdisplaybreaks

\begin{document}
\title{Compressive Periodogram Reconstruction\\Using Uniform Binning} 
\author{Dyonisius Dony Ariananda,~\IEEEmembership{Student Member,~IEEE}, Daniel Romero,~\IEEEmembership{Student Member,~IEEE}, \\ and~Geert Leus,~\IEEEmembership{Fellow,~IEEE}\vspace{-4ex}
\thanks{D.D. Ariananda and G. Leus are supported by NWO-STW under the VICI program (project 10382). They 
are with the Faculty of EEMCS, Delft University of Technology, Mekelweg 4, 2628 CD Delft, The Netherlands, email: \{d.a.dyonisius, g.j.t.leus\}@tudelft.nl. D. Romero is supported by ERDF, TEC2010-21245-C02-02/TCM DYNACS, CONSOLIDERINGENIO2010 CSD2008-00010 COMONSENS, FPU Grant AP2010-0149 and CN 2012/260 AtlantTIC. He is 
with Dept. of Signal Theory and Communications, University of Vigo, Spain, email: dromero@gts.uvigo.es. Part of this work was presented in~\cite{CAMSAP13}.}} 

\maketitle
\thispagestyle{empty}
\pagenumbering{arabic}
{\color{blue}\begin{abstract}
In this paper, two problems that show great similarities are examined. The first problem is the reconstruction of the angular-domain periodogram from spatial-domain signals received at different time indices. The second one is the reconstruction of the frequency-domain periodogram from time-domain signals received at different wireless sensors. We split the entire angular or frequency band into uniform bins. The bin size is set such that the received spectra at two frequencies or angles, whose distance is equal to or larger than the size of a bin, are uncorrelated. These problems in the two different domains lead to a similar circulant structure in the so-called coset correlation matrix. 
This circulant structure allows for a strong compression and a simple least-squares reconstruction method. The latter is possible under the full column rank condition of the system matrix, which can be achieved by designing the spatial or temporal sampling patterns based on a circular sparse ruler. 
We analyze the statistical performance of the compressively reconstructed periodogram including bias and variance. 
We further consider the case when the bins are so small 
that the received spectra at two frequencies or angles, with a spacing between them larger than the size of the bin, can still be correlated. In this case, the resulting coset correlation matrix is generally not circulant and thus a special approach is required. 
\end{abstract}}


\section{Introduction}\label{introduction}

The similarity between 
spectral analysis problems in the spatial-angular domain and 
in the time-frequency domain has attracted signal processing researchers since the 
1970s. 
Direction of arrival (DOA) estimation and frequency identification of sinusoids are examples of such similar 
problems examined during that period~\cite{Stoica}. 
The renewed interest in 
spectral analysis problems, especially due to the emergence of compressive sampling, has spurred reinvestigations on this similarity because, when 
time-domain or spatial-domain compression is introduced, this similarity 
can be 
exploited to tackle different problems using the same algorithmic approach. 

This paper focuses on both the reconstruction of the angular-domain periodogram 
from far-field signals received by an antenna array at different time indices (problem P1) and that of the 
frequency-domain periodogram from the time-domain signals 
received by different wireless sensors (problem P2). It further underlines the similarity between P1 and P2. 
Unless 
otherwise stated, the entire angular or frequency band is divided into uniform 
bins, where the size of the bins is configured such that 
the received spectra at two frequencies or angles, whose distance is equal to or larger than the size of a bin, are uncorrelated. 
In this case, 
the so-called coset correlation matrix will have a circulant structure, which allows the use of a periodic non-uniform linear array (non-ULA) in P1 and a multi-coset sampler in P2 in order to produce a strong compression.

Our work in P1 is motivated in part by~\cite{Kochman}, 
which attempts to reconstruct 
the angular spectrum from spatial-domain samples received by a non-ULA. Comparable works to~\cite{Kochman} for P2 are~\cite{Venkataramani} and~\cite{Eldar}, which focus on the analog signal reconstruction from its sub-Nyquist rate samples. However, the aim of~\cite{Kochman}-\cite{Eldar} to reconstruct 
the original spectrum or signal leads to an underdetermined problem, which has a unique solution only if we add 
constraints on the spectrum such as a sparsity constraint. A less ambitious goal in the context of P2 is to 
reconstruct the power spectrum instead of the actual signal from sub-Nyquist rate samples. 
For wide-sense stationary (WSS) signals, this has been shown to be possible in~\cite{Lexa} and~\cite{TSP12} without applying a sparsity constraint on the power spectrum. 
Meanwhile, the work of~\cite{XiaodongWang} 
assumes the existence of a multiband signal where 
different bands are uncorrelated. In this case, the diagonal structure of the correlation matrix of the entries at different bands can be exploited. 
Note though that~\cite{XiaodongWang} does not focus on the strongest compression rate and 
uses frequency smoothing to approximate the correlation matrix computation as it 
relies on a single 
realization of the received signal. 
Comparable works to~\cite{TSP12} in P1 are~\cite{Nested}-\cite{Siavash}, which aim to estimate the DOA of uncorrelated point sources with fewer 
antennas than sources. 
This is possible because for uncorrelated point sources, 
the spatial correlation matrix of the received signals also has a Toeplitz structure. Hence, for a given ULA, 
we can deactivate some 
antennas but still manage to estimate 
the spatial correlation 
at all lags. 
For example,~\cite{Nested} and~\cite{Coprime} suggest to place 
the active antennas based on a nested or coprime array, respectively, 
which results in a longer virtual array called the difference co-array (which is uniform in this case). As the difference co-array generally has more antennas and a larger aperture than the actual array, the degrees of freedom are increased allowing~\cite{Nested} and~\cite{Coprime} to estimate the DOA of more uncorrelated sources than sensors. 
In a more optimal way, a uniform difference co-array can also be obtained by the minimum redundancy array (MRA) of~\cite{Moffet}, 
but the nested and coprime arrays present many advantages 
due to their algebraic construction. 
MRAs have been used in~\cite{Siavash} to estimate the DOA of more uncorrelated sources than sensors, or more generally, to estimate the angular-domain power spectrum. 

Unlike~\cite{Kochman}, our work for P1 
focuses on the angular periodogram reconstruction (similar to~\cite{Siavash}). This allows us to have an overdetermined problem that is solvable even without a sparsity constraint on the angular domain. This is beneficial for applications that require only information about the angular periodogram and not the actual angular spectrum. 
Our work is also different from~\cite{Nested}-\cite{Siavash} as we do not exploit the Toeplitz structure of the spatial correlation matrix. 
As for P2, 
we focus on frequency periodogram reconstruction (unlike~\cite{Venkataramani}-\cite{Eldar}) but we do not exploit the Toeplitz structure of the time-domain correlation matrix 
(unlike~\cite{TSP12}). On the other hand, the problem handled by~\cite{XiaodongWang} can be considered as a special case of P2 but, unlike~\cite{XiaodongWang}, we aim for the strongest compression rate which is achieved by exploiting the circulant structure of the coset correlation matrix and 
solving the minimal circular sparse ruler problem. 
Moreover, unlike~\cite{XiaodongWang}, we also exploit the signals received by different sensors to estimate 
the correlation matrix. 

Also related to P2, a cooperative compressive wideband spectrum sensing scheme for cognitive radio (CR) networks is proposed 
in~\cite{FZheng}.
While 
\cite{FZheng} can reduce the required sampling rate per CR, its focus on reconstructing the spectrum or the spectrum support requires a sparsity constraint on the original spectrum. 
Unlike~\cite{FZheng},~\cite{AsilomarCR} 
focuses on compressively estimating the power spectrum instead of the spectrum 
by extending~\cite{TSP12} 
for a cooperative scenario. 
However, while 
the required sampling rate per sensor can be lowered without applying a sparsity constraint on the power spectrum, the exploitation of the cross-spectra between signals 
at different sensors in~\cite{AsilomarCR} 
requires the knowledge of the channel state information (CSI). 
Our approach for P2 does not require a sparsity constraint on the original periodogram (unlike~\cite{FZheng}) and it does not require CSI since 
we are not interested in the cross-spectra between samples at different sensors (unlike~\cite{AsilomarCR}). 
In~\cite{FrugalSensing}, each wireless sensor applies a threshold on the measured average signal power after applying a random wideband filter. 
The threshold output is then communicated as a few bits to a fusion centre, which uses them 
to recover the power spectrum by generalizing the problem in the form of inequalities. 
The achievable compression rate with such a system is not clear though, in contrast to what we will present in this paper. 

In more advanced problems, such as cyclic spectrum reconstruction from sub-Nyquist rate samples of cyclostationary signals in~\cite{GerryCyclo}-\cite{GeertCyclo}
or angular power spectrum reconstruction from signals produced by correlated sources in~\cite{ElsevierDOA}, 
finding a special structure in the resulting correlation matrix that can be exploited to perform compression is 
challenging. 
A similar challenge is faced in Section~\ref{correlatedbins}, where we 
consider the case when we reduce the 
bin size 
such that the received spectra at two frequencies or angles with a spacing 
larger than the bin size can still be correlated. 
As the resulting coset correlation matrix in this case is generally not circulant, 
we further develop the concepts originally introduced in~\cite{GeertCyclo} 
and~\cite{ElsevierDOA} 
to solve our problem.

{\color{blue}We now would like to summarize the advantages of our approach and highlight our contribution.
\begin{itemize}
\item We propose a compressive periodogram reconstruction approach, which does not rely on 
any sparsity constraint on the original signal or the periodogram. Moreover, it is based on a simple least-squares (LS) algorithm leading to a low complexity. 
\item In our approach, we also focus on the strongest possible compression that maintains the identifiability of the periodogram, which is shown 
 to be related to a minimal circular sparse ruler. 
\item 
Our approach does not require any knowledge of the CSI. \item The statistical performance analysis of the compressively reconstructed periodogram is also provided.
\item Our approach can also be modified to handle cases where the spectra in different bins are correlated. 
\end{itemize}
This paper is organized as follows. The system model description (including the definition of the so-called coset correlation matrix) and the problem statement are provided in Section~\ref{uncorr_bins_system_model}. Section~\ref{compression_reconstruction} discusses the spatial (for P1) or temporal (for P2) compression as well as 
periodogram reconstruction 
using LS. Here, the condition for the system matrix to have full column rank and its connection to the minimal circular sparse ruler problem are 
provided. 
Section~\ref{corr_mat_approximate} shows how to approximate the expectation operation in the correlation matrix computation and summarizes the procedure to compressively estimate the periodogram. In Section~\ref{performance}, we provide an analysis on the statistical performance of the compressively reconstructed periodogram 
including a bias and variance analysis. 
Sections~\ref{uncorr_bins_system_model}-\ref{performance} assume that the received signals at different time instants (for P1) or at different sensors (for P2) have the same statistics. 
To handle more general cases, we propose a multi-cluster model in Section~\ref{CaseC2}, which considers clusters of time indices in P1 or clusters of sensors in P2 and assumes that the signal 
statistics are only constant within a cluster. Another 
case is discussed in Section~\ref{correlatedbins}, where 
the received spectra at two frequencies or angles located at different predefined bins can still be correlated. Some numerical studies are elaborated in Section~\ref{numerical} and Section~\ref{sec:conclusion} provides conclusions.

{\it Notation:} Upper (lower) boldface letters are used to denote matrices (column vectors). Given an $N \times N$ matrix ${\bf X}$, diag$({\bf X})$ is an $N\times 1$ vector containing the main diagonal entries of ${\bf X}$. Given an $N \times 1$ vector ${\bf x}$, diag$({\bf x})$ is an $N\times N$ diagonal matrix whose diagonal entries are given by the entries of ${\bf x}$.}

\section{System Model
}\label{uncorr_bins_system_model} 
\subsection{Model Description and Problem Statement}\label{model_problem_statement}

We aim at estimating the following spectral representation of the power 
of a 
process $x[\tilde{n}]$: 
\vspace{-0.5mm}
\begin{eqnarray}
P_x(\vartheta)&=&\lim_{\tilde{N}\rightarrow\infty}E\left\{\frac{1}{\tilde{N}}\left|\sum_{\tilde{n}=0}^{\tilde{N}-1}x[\tilde{n}]e^{-j\vartheta\tilde{n}}\right|^2\right\}\nonumber\\
&=&\lim_{\tilde{N}\rightarrow\infty}E\left\{\frac{1}{\tilde{N}}\left|X_{(\tilde{N})}(\vartheta)\right|^2\right\}.
\label{eq:PowerSpectrum}
\vspace{-0.5mm}
\end{eqnarray}
Here, $x[\tilde{n}]$ represents either the spatial-domain process at the output of a ULA for P1 or the time-domain process sensed by a wireless sensor for P2. In addition, $X_{(\tilde{N})}(\vartheta)$ represents either the value of the angular spectrum at angle $\text{sin}^{-1}(2\vartheta)$ for P1 or that of the frequency spectrum at frequency $\vartheta$ for P2, with $\vartheta \in [-0.5,0.5)$. 
Note from~\cite{Stoica} that, for a {\it WSS process} $x[\tilde{n}]$, $P_x(\vartheta)$ represents the {\it power spectrum}. 
To estimate $P_x(\vartheta)$ 
in~\eqref{eq:PowerSpectrum}, consider the $\tilde{N} \times 1$ complex-valued observation vectors ${\bf x}_t=[x_t[0],x_t[1],\dots,x_t[{\tilde{N}-1}]]^T$, $t=1,2\dots,\tau$, where $x_t[\tilde{n}]$ represents the output of the $(\tilde{n}+1)$-th antenna in the ULA of $\tilde{N}$ half-wavelength spaced antennas at time index $t$ for P1
or the $(\tilde{n}+1)$-th sample out of $\tilde{N}$ successive 
samples produced by the 
Nyquist-rate sampler at the $t$-th sensor for P2. 
To acquire an accurate 
Fourier interpretation, we assume a relatively large $\tilde{N}$, which is affordable for P2 and also realistic for P1, 
if we consider millimeter wave imaging applications where the antenna spacing is very small and thus the required aperture has to be covered by a large number of antennas~\cite{Kochman}. 
Denote the discrete-time Fourier transform (DTFT) of $x_t[\tilde{n}]$ by $X_t(\vartheta)$.
As $X_t(\vartheta)$ at $\vartheta \in [-0.5,0)$ is a replica of $X_t(\vartheta)$ at $\vartheta \in [0.5,1)$, we can focus on $X_t(\vartheta)$ in $\vartheta \in [0,1)$. 

Next, we divide the $\tilde{N}$ uniform grid points (that is, 
the antennas of the ULA for P1 or the indices of the Nyquist-rate samples for P2) into $L$ non-overlapping blocks of $N$ uniform grid points. 
We collect all the $(n+1)$-th grid points from each of the $L$ blocks and label this collection of grid points, i.e., $\{\tilde{n} \in \{0,1,\dots,\tilde{N}-1\}|\tilde{n}\text{ mod }N=n\}$, as the {\it $(n+1)$-th coset}, with $\tilde{n} \text{ mod }N$ the remainder of the integer division $\tilde{n}/N$. In this paper, the {\it coset index} of the $(n+1)$-th coset is $n$. This procedure allows us to view the above uniform sampling as a multi-coset sampling~\cite{Venkataramani} with $N$ cosets. Consequently, the ULA of $\tilde{N}$ antennas in P1 can be regarded as $N$ interleaved uniform linear subarrays (ULSs)~\cite{Kochman} (which are the cosets) of $L$ $(N\lambda/2)$-spaced antennas 
with $\lambda$ the wavelength, whereas the 
$\tilde{N}$ time-domain samples in P2 can be considered as the output of a time-domain multi-coset sampler with $L$ samples per coset. If we activate only the $(n+1)$-th coset, the spatial- or time-domain samples at index $\tilde{n}$ are given by
\vspace{-0.5mm}
\begin{equation}
\bar{x}_{t,n}[\tilde{n}]=x_t[\tilde{n}]\sum_{l=0}^{L-1}\delta[\tilde{n}-(lN+n)], \: n=0,1,\dots,N-1, 
\label{eq:x_{t,n}}
\vspace{-0.5mm}
\end{equation}
which can be collected into the $\tilde{N} \times 1$ vector $\bar{\bf x}_{t,n}=[\bar{x}_{t,n}[0],$ $\bar{x}_{t,n}[1],\dots,\bar{x}_{t,n}[{\tilde{N}-1}]]^T$. 
Observe that ${\bf x}_{t}=\sum_{n=0}^{N-1}\bar{\bf x}_{t,n}$. To show the relationship between the DTFT of $\bar{x}_{t,n}[\tilde{n}]$ and that of $x_t[\tilde{n}]$, 
we split $\vartheta \in [0,1)$ into $N$ equal-width bins and express the spectrum at the $(i+1)$-th bin ($i=0,1,\dots,N-1$) as 
$X_{t,i}(\vartheta)=X_t\left(\vartheta+\frac{i}{N}\right)$ with $\vartheta$ now limited to $\vartheta \in [0,1/N)$. 
As either the spatial or temporal sampling rate becomes $1/N$ times the Nyquist-rate when only the $(n+1)$-th coset is activated, the DTFT of $\bar{x}_{t,n}[\tilde{n}]$, denoted by $\bar{X}_{t,n}(\vartheta)$, is the sum of $N$ aliased versions of $X_t(\vartheta)$ at $N$ different bins.
This is shown 
for $n=0,1,\dots,N-1$ as~\cite{Eldar} 
\vspace{-1mm}
\begin{equation}
\bar{X}_{t,n}(\vartheta)=\frac{1}{N}\sum_{i=0}^{N-1}
X_{t,i}(\vartheta)e^{\frac{j2\pi n i}{N}},\quad \vartheta \in [0,1/N). 
\label{eq:X_{t,n}}
\vspace{-1mm}
\end{equation}
Collecting $\bar{X}_{t,n}(\vartheta)$, 
for $n=0,1,\dots,N-1$, into the $N\times 1$ vector $\bar{\bf x}_{t}(\vartheta)=[\bar{X}_{t,0}(\vartheta),\bar{X}_{t,1}(\vartheta),\dots,\bar{X}_{t,N-1}(\vartheta)]^T$ and introducing the $N\times 1$ vector ${\bf x}_{t}(\vartheta)=[X_{t,0}(\vartheta),X_{t,1}(\vartheta),$ $\dots,X_{t,N-1}(\vartheta)]^T$ allow us to write 
\vspace{-0.5mm}
\begin{equation}
\bar{\bf x}_{t}(\vartheta)={\bf B}{\bf x}_{t}(\vartheta),\quad \vartheta \in [0,1/N),
\label{eq:x_{t}_bar}
\vspace{-0.5mm}
\end{equation}   
with the element of the $N\times N$ matrix ${\bf B}$ at the $(n+1)$-th row and the $(i+1)$-th column given by $[{\bf B}]_{n+1,i+1}=\frac{1}{N}e^{\frac{j2\pi n i}{N}}$. 

\begin{figure}[h]
				\centering
        \includegraphics[width=0.4\textwidth]{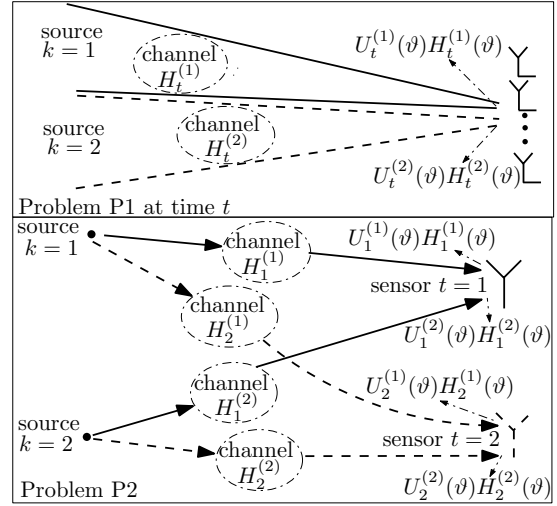}
        \caption{The system model for problems P1 and P2.}        
        \label{fig:SystemModel}
\end{figure}
We now assume the presence of $K$ active users, 
consider the model in Fig.~\ref{fig:SystemModel}, and introduce the following definition (see also Fig.~\ref{fig:SystemModel}).
\vspace{0.5mm}
\newline
\hspace*{1mm}{\it Definition 1: We define the complex-valued zero-mean random processes 
$U_t^{(k)}(\vartheta)$ and $H_t^{(k)}(\vartheta)$ as
\begin{itemize}
\item For P1, $U_t^{(k)}(\vartheta)$ is the source signal related to the $k$-th user received at time index $t$, which can depend on the DOA $\text{sin}^{-1}(2\vartheta)$ 
due to scattering. For P2, it is the source signal related to the $k$-th user received at sensor $t$, which can vary with frequency $\vartheta$ 
due to power loading,
\item $H_t^{(k)}(\vartheta)$ is the related channel response for the $k$-th user at time index $t$ and DOA $\text{sin}^{-1}(2\vartheta)$ (for P1) or at sensor $t$ and frequency $\vartheta$ (for P2). 
\end{itemize}}
\noindent Note from Fig.~\ref{fig:SystemModel} that, theoretically, $U_t^{(k)}(\vartheta)$ is the only component observed by the ULA in P1 or by the sensors in P2 if no fading channel exists.
Define $N_t(\vartheta)$ as the zero-mean additive white (both in $\vartheta$ and $t$) 
noise at DOA $\text{sin}^{-1}(2\vartheta)$ and time index $t$ (for P1) or at frequency $\vartheta$ and sensor $t$ (for P2). 
By introducing $N_{t,i}(\vartheta)=N_t\left(\vartheta+\frac{i}{N}\right)$ and similarly also $H_{t,i}^{(k)}(\vartheta)$ 
as well as $U_{t,i}^{(k)}(\vartheta)$, 
we can then use Definition~1 
to write $X_{t,i}(\vartheta)$ in~\eqref{eq:X_{t,n}} as
\vspace{-0.5mm}
\begin{equation}
X_{t,i}(\vartheta)=\sum_{k=1}^{K}H_{t,i}^{(k)}(\vartheta) U_{t,i}^{(k)}(\vartheta)+N_{t,i}(\vartheta),\: \vartheta \in [0,1/N).
\label{eq:Xti_as_H_tki_and_Uki}
\vspace{-0.5mm}
\end{equation}

Next, let us consider the following assumption. 
\vspace{0.5mm}
\newline
\hspace*{1mm}{\it Assumption~1: 
$X_{t,i}(\vartheta)$ in~\eqref{eq:Xti_as_H_tki_and_Uki} is an ergodic stochastic process along 
$t$. 
}
\vspace{0.5mm}
\newline
This ergodicity assumption requires that the statistics of ${\bf x}_{t}(\vartheta)$ in~\eqref{eq:x_{t}_bar} do not change with $t$ (a more general case is discussed in Section~\ref{CaseC2}). Hence, we can define the $N\times N$ correlation matrix of ${\bf x}_{t}(\vartheta)$ as ${\bf R}_x(\vartheta)=E[{\bf x}_{t}(\vartheta){\bf x}_{t}^H(\vartheta)]$, for all $t$ and $\vartheta \in [0,1/N)$.
The assumption that the statistics of ${\bf x}_{t}(\vartheta)$ 
do not vary with $t$ is motivated for P1 when the signal received by the array is stationary in the time-domain. For P2, it implies that the statistics of the signal ${\bf x}_t$ received by different sensors $t$ are the same. Observe from~\eqref{eq:Xti_as_H_tki_and_Uki} that the element of ${\bf R}_x(\vartheta)$ at the $(i+1)$-th row and the $(i'+1)$-th column is given by
\vspace{-0.5mm}
\begin{align}
&E[X_{t,i}(\vartheta){X_{t,i'}^*}(\vartheta)]=
E[|N_{t,i}(\vartheta)|^2]\delta[i-i']+\nonumber \\
&\sum_{k=1}^{K}\sum_{k'=1}^{K}E[U_{t,i}^{(k)}(\vartheta){U_{t,i'}^{(k')*}}(\vartheta)]E[H_{t,i}^{(k)}(\vartheta){H_{t,{i'}}^{(k')*}}(\vartheta)], 
\label{eq:E[XtXt']}
\vspace{-0.5mm}
\end{align}
where we assume that the source signal $U_t^{(k)}(\vartheta)$, the noise $N_t(\vartheta)$, and the channel response $H_t^{(k)}(\vartheta)$ are mutually uncorrelated. We now consider the following remark.
\vspace{0.5mm}
\newline
\hspace*{1mm}{\it Remark~1: 
The diagonal of ${\bf R}_x(\vartheta)$, which is given by $\{E[|X_{t,i}(\vartheta)|^2]\}_{i=0}^{N-1}$ 
and which is independent of $t$, 
can be related to 
$P_x(\vartheta)$ in~\eqref{eq:PowerSpectrum}. In practice, this expected value has to be estimated and 
Assumption~1 allows us to estimate $E[|X_{t,i}(\vartheta)|^2]$ using $\frac{1}{\tau}\sum_{t=1}^{\tau}|X_{t,i}(\vartheta)|^2$. We can then consider $\frac{1}{\tilde{N}\tau}\sum_{t=1}^{\tau}|X_{t,i}(\vartheta)|^2$ as a reasonable estimate for 
$P_x(\vartheta+\frac{i}{N})$ in~\eqref{eq:PowerSpectrum}, for $\vartheta \in [0,1/N)$. Here, $\frac{1}{\tilde{N}\tau}\sum_{t=1}^{\tau}|X_t(\vartheta)|^2$, for $\vartheta \in [0,1)$, can be considered as the averaged periodogram (AP) of $x_t[\tilde{n}]$ over different time indices $t$ in P1 or different sensors $t$ in P2.} 
\vspace{0.5mm}
\newline
Note that, even for the noiseless case, we can expect $X_{t,i}(\vartheta)$ in~\eqref{eq:Xti_as_H_tki_and_Uki} 
to vary with $t$ if either one (or both) of the following situations occurs. 
\begin{list}{\labelitemi}{\leftmargin=0mm \itemindent=0.5em}
\item For P1, $U_t^{(k)}(\vartheta)$ varies with the time index $t$ if the information that is being transmitted changes with time.
For P2, it varies with the sensor index $t$ where the signal is received if the sensors are not synchronized. 
\item For P1, $H_t^{(k)}(\vartheta)$ varies with the time index $t$ if Doppler fading effects exist.
For P2, it varies with the sensor index $t$ where the signal is received, due to path loss, shadowing, and small-scale spatial fading effects. 
\end{list}
We then consider the following remark.
\vspace{0.5mm}
\newline
\hspace*{1mm}{\it Remark~2: Recall that the size of the predefined bins in $\vartheta \in [0, 1)$ is a design parameter given by $\frac{1}{N}$, i.e., the inverse of the number of cosets. 
Using~\eqref{eq:E[XtXt']}, it is easy to find that ${\bf R}_{x}(\vartheta)$ is a diagonal matrix if either
$E[U_t^{(k)}(\vartheta)U_t^{(k')*}(\vartheta')]=0$ and/or $E[H_t^{(k)}(\vartheta)H_t^{(k')*}(\vartheta')]=0$ for $|\vartheta'-\vartheta|\geq \frac{1}{N}$, with $\vartheta,\vartheta' \in [0,1)$, and for all $t,k,k'$.} 
\vspace{0.5mm}
\newline
One example 
for both P1 and P2 is when we have $K$ non-overlapping active bands corresponding to $K$ different users leading to a multiband structure in the $\vartheta$-domain with either the $K$ different users transmitting mutually uncorrelated source signals and/or the signals from the $K$ different users passing through mutually uncorrelated wireless channels on their way to the receiver. If we denote the support of the $k$-th active band by ${\mathcal{B}_k}$ and its 
bandwidth by $\Lambda({\mathcal{B}_k})=\text{sup}\{{\mathcal{B}_k}\}-\text{inf}\{{\mathcal{B}_k}\}$, the condition in Remark~2 is then satisfied by setting $N$ such that $\frac{1}{N}\geq \max_k\Lambda({\mathcal{B}}_k)$. Note that such a choice is possible, 
especially for P2, as the channelization parameter for a communication network is usually known. 

We focus on the case where ${\bf R}_{x}(\vartheta)$ is a diagonal matrix and 
define the so-called $N\times N$ {\it coset correlation matrix} as
\begin{equation}
{\bf R}_{\bar{x}}(\vartheta)=E[\bar{\bf x}_t(\vartheta)\bar{\bf x}_t^H(\vartheta)]
={\bf B}{\bf R}_{x}(\vartheta){\bf B}^H,\:\: \vartheta \in [0,1/N).
\label{eq:Rxt_bar}
\end{equation}
Observe that ${\bf R}_{\bar{x}}(\vartheta)$ is a circulant matrix when ${\bf R}_{x}(\vartheta)$ is a diagonal matrix since ${\bf B}$ is an inverse discrete Fourier transform (IDFT) matrix, as can be concluded from~\eqref{eq:x_{t}_bar}. 
Based on the aforementioned system model, we finally formulate our problem statement as follows:
\vspace{0.5mm}
\newline
\hspace*{1mm}{\it Problem Statement: As an estimate of the 
spectral representation of the power $P_x(\vartheta)$ in~\eqref{eq:PowerSpectrum} (which is also the power spectrum when $x[\tilde{n}]$ in~\eqref{eq:PowerSpectrum} is a WSS process), we aim to compressively reconstruct the AP of $x_t[\tilde{n}]$ in~\eqref{eq:x_{t,n}} over the index $t$, where we assume that $x_t[\tilde{n}]$ is ergodic along the index t and that its coset correlation matrix ${\bf R}_{\bar{x}}(\vartheta)$ has a circulant structure. We discuss the compression and the reconstruction in Section~\ref{compression_reconstruction} and the estimation of the correlation matrix in Section~\ref{corr_mat_approximate}.}

\subsection{Interpretation of AP in Remark~1}

How the AP in Remark~1 is interpreted with respect to $U_t^{(k)}(\vartheta)$ and $H_t^{(k)}(\vartheta)$ depends on which of the functions varies in $t$. 
For example, 
consider problem~P2 and assume that only one user $k$ can occupy a given frequency $\vartheta$ at a given time 
and that only $H_t^{(k)}(\vartheta)$ varies in $t$, i.e., $U_t^{(k)}(\vartheta)=U^{(k)}(\vartheta)$. 
For this example,
we have from~\eqref{eq:Xti_as_H_tki_and_Uki}
\begin{align}
&\frac{1}{\tilde{N}\tau}\sum_{t=1}^{\tau}|X_t(\vartheta)|^2
=\frac{|U^{(k)}(\vartheta)|^2}{\tilde{N}}\sum_{t=1}^{\tau}\frac{|H_t^{(k)}(\vartheta)|^2}{\tau}\nonumber \\
&+\sum_{t=1}^{\tau}\frac{|N_t(\vartheta)|^2}{\tilde{N}\tau}
+\sum_{t=1}^{\tau}\frac{2\text{Re}(H_t^{(k)}(\vartheta)U^{(k)}(\vartheta)N^*_t(\vartheta))}
{\tilde{N}\tau},
\label{ExS3P2}
\end{align}
where $\text{Re}(x)$ gives the real component of $x$, 
the first term is the 
classical periodogram of the user signals $\frac{|U^{(k)}(\vartheta)|^2}{\tilde{N}}$ scaled by the averaged fading magnitude experienced at different channels $\frac{1}{\tau}\sum_{t=1}^{\tau}|H_t^{(k)}(\vartheta)|^2$, the second term is the AP of the noises at different sensors $t$, and the 
last term 
converges to zero as $\tau$ becomes 
larger 
due to the uncorrelatedness between the noise $N_t(\vartheta)$ and the channel response $H_t^{(k)}(\vartheta)$. The 
assumption that the statistics of $X_t(\vartheta)$ do not vary with 
$t$ (as required by Assumption~1) 
implies that the statistics of the fading experienced by different sensors $t$ are the same 
(e.g., they experience small-scale 
fading on top of the same path loss and shadowing).

As another example, consider problem P1 
and assume that only one user $k$ can occupy a given DOA $\text{sin}^{-1}(2\vartheta)$ at a given time 
and that only $U_t^{(k)}(\vartheta)$ varies in $t$, i.e., $H_t^{(k)}(\vartheta)=H^{(k)}(\vartheta)$.
For this example, we have from~\eqref{eq:Xti_as_H_tki_and_Uki}
\begin{align}
&\frac{1}{\tilde{N}\tau}\sum_{t=1}^{\tau}|X_t(\vartheta)|^2
=|H^{(k)}(\vartheta)|^2\sum_{t=1}^{\tau}\frac{|U_t^{(k)}(\vartheta)|^2}{\tilde{N}\tau}\nonumber\\
&+\sum_{t=1}^{\tau}\frac{|N_t(\vartheta)|^2}{\tilde{N}\tau}
+\sum_{t=1}^{\tau}\frac{2\text{Re}(U_t^{(k)}(\vartheta)H^{(k)}(\vartheta)N^*_t(\vartheta))}
{\tilde{N}\tau},
\label{ExS1P1}
\end{align}
where the first term is the angular-domain 
AP of the user signals $\frac{1}{\tilde{N}\tau}\sum_{t=1}^{\tau}|U_t^{(k)}(\vartheta)|^2$ scaled by the magnitude of the time-invariant channel angular response $|H^{(k)}(\vartheta)|^2$, the second term is the angular-domain AP 
of the noise, 
and the last term again 
converges to zero as $\tau$ becomes 
larger due to the uncorrelatedness between $N_t(\vartheta)$ and $U_t^{(k)}(\vartheta)$.

\section{Compression and Reconstruction}\label{compression_reconstruction}
\subsection{Spatial or Temporal Compression}\label{uncorr_bins_compression}
As ${\bf R}_{\bar{x}}(\vartheta)$ in~\eqref{eq:Rxt_bar} is a circulant matrix, 
it is possible to 
condense its entries 
into an $N\times 1$ vector ${\bf r}_{\bar{x}}(\vartheta)=[{r}_{\bar{x}}(\vartheta,0),$ ${r}_{\bar{x}}(\vartheta,1),\dots,{r}_{\bar{x}}(\vartheta,N-1)]^T$
with ${r}_{\bar{x}}(\vartheta,(n-n')\text{ mod }N)=E\left[\bar{X}_{t,n}(\vartheta),\bar{X}^*_{t,n'}(\vartheta)\right]$. 
We can then relate ${\bf r}_{\bar{x}}(\vartheta)$ to ${\bf R}_{\bar{x}}(\vartheta)$ as
\begin{equation}
\text{vec}({\bf R}_{\bar{x}}(\vartheta))={\bf T}{\bf r}_{\bar{x}}(\vartheta),\quad \vartheta \in [0,1/N),
\label{eq:Rbarx_as_rbar_x}
\end{equation}
where ${\bf T}$ is an $N^2\times N$ repetition matrix whose $(q+1)$-th row is given by the $\left(\left(q-\left\lfloor\frac{q}{N}\right\rfloor\right)\text{ mod }N+1\right)$-th row of the $N\times N$ identity matrix ${\bf I}_N$ and vec$(.)$ is the operator that stacks all columns of a matrix into one column vector.
The possibility to condense the $N^2$ entries of ${\bf R}_{\bar{x}}(\vartheta)$ into the $N$ entries of ${\bf r}_{\bar{x}}(\vartheta)$ facilitates compression 
by performing a spatial- or time-domain non-uniform periodic sampling (similar to~\cite{Eldar}), in which only 
$M<N$ cosets are activated. 
Here, we use the set $\mathcal{M}=\{n_0, n_1,\dots, n_{M-1}\}$, with $0\leq n_0 < n_1 < \dots < n_{M-1}\leq N-1$, to indicate the indices of the $M$ active cosets. 
All values of $\bar{x}_{t,n}[\tilde{n}]$ in~\eqref{eq:x_{t,n}} are then collected and their corresponding DTFT $\bar{X}_{t,n}(\vartheta)$ in~\eqref{eq:X_{t,n}} is computed for all $n \in \mathcal{M}$. 
Stacking $\left\{\bar{X}_{t,n}(\vartheta)\right\}_{n\in\mathcal{M}}$ into the $M\times 1$ vector $\bar{\bf y}_{t}(\vartheta)=[\bar{X}_{t,n_0}(\vartheta),\bar{X}_{t,n_1}(\vartheta),\dots,\bar{X}_{t,n_{M-1}}(\vartheta)]^T$ allows us to relate 
$\bar{\bf y}_{t}(\vartheta)$ to $\bar{\bf x}_{t}(\vartheta)$ in~\eqref{eq:x_{t}_bar} as
\begin{equation}
\bar{\bf y}_{t}(\vartheta)={\bf C}\bar{\bf x}_{t}(\vartheta),\quad \vartheta \in [0,1/N),
\label{eq:y_{t}_bar}
\end{equation}  
where ${\bf C}$ is an $M \times N$ selection matrix whose rows are selected from the rows of ${\bf I}_N$ based on ${\mathcal{M}}$. 
Since ${\bf C}$ is real, the $M\times M$ correlation matrix of $\bar{\bf y}_{t}(\vartheta)$, for $\vartheta \in [0,1/N)$, 
can be written as
\begin{equation}
{\bf R}_{\bar{y}}(\vartheta)=E[\bar{\bf y}_{t}(\vartheta)\bar{\bf y}^H_{t}(\vartheta)]={\bf C}{\bf R}_{\bar{x}}(\vartheta){\bf C}^T. 
\label{eq:Ry_bar}
\end{equation}
We then take $\eqref{eq:Rbarx_as_rbar_x}$ into account, cascade all columns of ${\bf R}_{\bar{y}}(\vartheta)$ into a column vector $\text{vec}({\bf R}_{\bar{y}}(\vartheta))$, and write
\begin{equation}
\text{vec}({\bf R}_{\bar{y}}(\vartheta))
={\bf R}_c{\bf r}_{\bar{x}}(\vartheta),\quad \vartheta \in [0,1/N),
\label{eq:Rybar_as_rbar_x}
\end{equation}
where ${\bf R}_c=({\bf C}\otimes{\bf C}){\bf T}$ is a real $M^2\times N$ matrix and $\otimes$ denotes the Kronecker product operation.

\subsection{Reconstruction}\label{uncorrbinsreconstruct}
If ${\bf R}_c$ in~\eqref{eq:Rybar_as_rbar_x} is a tall matrix ($M^2 \geq N$), which is possible despite $M <N$, and if it has full column rank, 
${\bf r}_{\bar{x}}(\vartheta)$ in~\eqref{eq:Rybar_as_rbar_x} can be reconstructed from $\text{vec}({\bf R}_{\bar{y}}(\vartheta))$ using 
LS for all $\vartheta \in [0,1/N)$. In addition, as long as the identifiability of ${\bf r}_{\bar{x}}(\vartheta)$ in~\eqref{eq:Rybar_as_rbar_x} is preserved, we can also consider estimators other than LS (such as in~\cite{Daniel}). To formulate a necessary and sufficient condition for the identifiability of ${\bf r}_{\bar{x}}(\vartheta)$ in~\eqref{eq:Rybar_as_rbar_x} from $\text{vec}({\bf R}_{\bar{y}}(\vartheta))$, let us review the concept of a circular sparse ruler defined in~\cite{Romero}. 
\vspace{0.5mm}
\newline
\hspace*{1mm}{\it Definition 2: A circular sparse ruler of length $N-1$ is a set $\mathcal{K} \subset \{0,1,\dots,N-1\}$ for which $\Omega(\mathcal{K})=\{(\kappa-\kappa')\text{ mod }N|\forall \kappa,\kappa' \in \mathcal{K}\}=\{0,1,\dots,N-1\}$. We call it minimal if there is 
no other circular sparse ruler of length $N-1$ with fewer elements.}
\newline
Detailed information about circular 
sparse rulers can be found in~\cite{Romero}. We can then use this concept 
to formulate the following theorem whose proof is available in~\cite{CAMSAP13}. 
\vspace{0.5mm}
{\color{blue}\newline
\hspace*{1mm}{\it Theorem 1: ${\bf r}_{\bar{x}}(\vartheta)$ in~\eqref{eq:Rybar_as_rbar_x} is identifiable from $\text{vec}({\bf R}_{\bar{y}}(\vartheta))$, i.e., ${\bf R}_c$ has full column rank, if and only if ${\mathcal M}$ is a circular sparse ruler, i.e., $\Omega(\mathcal{M}) = \{ 0,1,\dots,N-1\}$. When this is satisfied, 
${\bf R}_c$ contains all rows of ${\bf I}_N$.} 
\newline
\hspace*{1mm}Our goal is to obtain the strongest possible compression rate $M/N$ preserving the identifiability.} 
This is achieved by minimizing the cardinality of the set $\mathcal{M}$, $|\mathcal{M}|=M$,
under the condition that $\Omega({\mathcal M})=\{0,1,\dots, N-1\}$. 
This leads to a length-$(N-1)$ minimal circular sparse ruler problem, which can be written as 
\vspace{-0.9mm}
\begin{equation}
\min_{\mathcal{M}}\left|\mathcal{M}\right| \: \text{s.t.} \: \Omega(\mathcal{M})=\left\{0,1,\dots,N-1\right\}.
\label{eq:cardinality_comp}
\vspace{-0.9mm}
\end{equation}
Solving~\eqref{eq:cardinality_comp} minimizes the compression rate $M/N$ while maintaining the identifiability of ${\bf r}_{\bar{x}}(\vartheta)$ in~\eqref{eq:Rybar_as_rbar_x}. 

Recall that, for P1, $\mathcal{M}$ indicates the indices of the $M<N$ active ULSs in our ULA, which will be referred to 
as the ${\it underlying}$ array.  Therefore, we have a periodic non-ULA of active antennas and $\mathcal{M}$ governs the location of the active antennas in each spatial period. When $\mathcal{M}$ is a solution of the minimal length-$(N-1)$ circular sparse ruler problem in~\eqref{eq:cardinality_comp}, we can label the resulting non-ULA of active antennas as a {\it periodic circular MRA} and each of its spatial periods as a {\it circular MRA}. Similarly for P2, we can label the non-uniform sampling in each temporal period as {\it minimal circular sparse ruler sampling} and the entire periodic non-uniform sampling as {\it periodic minimal circular sparse ruler sampling} if the indices of the $M<N$ active cosets 
are given by the solution of~\eqref{eq:cardinality_comp}.

Once ${\bf r}_{\bar{x}}(\vartheta)$ is reconstructed from $\text{vec}({\bf R}_{\bar{y}}(\vartheta))$ in~\eqref{eq:Rybar_as_rbar_x} using LS for $\vartheta \in [0,1/N)$, 
we can use~\eqref{eq:Rbarx_as_rbar_x} to compute ${\bf R}_{\bar{x}}(\vartheta)$ from ${\bf r}_{\bar{x}}(\vartheta)$ and~\eqref{eq:Rxt_bar} to compute ${\bf R}_{x}(\vartheta)$ from ${\bf R}_{\bar{x}}(\vartheta)$ as ${\bf R}_{x}(\vartheta)=N^2{\bf B}^{H}{\bf R}_{\bar{x}}(\vartheta){\bf B}$. As 
we have $\text{diag}({\bf R}_{x}(\vartheta))=[E[|X_{t,0}(\vartheta)|^2],E[|X_{t,1}(\vartheta)|^2],\dots,E[|X_{t,N-1}(\vartheta)|^2]]^T$ with 
$\vartheta \in [0,1/N)$, reconstructing $\text{diag}({\bf R}_{x}(\vartheta))$ for all $\vartheta \in [0,1/N)$ gives $E[|X_t(\vartheta)|^2]$ for all $\vartheta \in [0,1)$. 

\section{Correlation Matrix Estimation}\label{corr_mat_approximate}

In practice, the expectation in~\eqref{eq:Ry_bar} must be approximated. 
Here, we propose to approximate the expectation in~\eqref{eq:Ry_bar} with the sample average 
over different time indices $t$ for P1 or sensors indices $t$ for P2, i.e., 
\vspace{-1mm}
\begin{equation}
\hat{\bf R}_{\bar{y}}(\vartheta)=\frac{1}{\tau}\sum_{t=1}^{\tau}\bar{\bf y}_{t}(\vartheta)\bar{\bf y}^H_{t}(\vartheta),\:\:\vartheta \in [0,1/N),
\label{eq:Rybar_hat}
\vspace{-1mm}
\end{equation}
where we recall that $\tau$ is either the total number of time indices or sensors from which the observations are collected. 
Observe that the $M\times M$ matrix $\hat{\bf R}_{\bar{y}}(\vartheta)$ 
is an unbiased estimate of ${\bf R}_{\bar{y}}(\vartheta)$ in~\eqref{eq:Rybar_as_rbar_x}. 
It is also a consistent estimate if 
Assumption~1 holds.
We can then apply 
LS reconstruction on $\hat{\bf R}_{\bar{y}}(\vartheta)$ in~\eqref{eq:Rybar_hat} instead of ${\bf R}_{\bar{y}}(\vartheta)$ in~\eqref{eq:Rybar_as_rbar_x}. 
As a result, the procedure to compressively reconstruct the AP of $x_t[\tilde{n}]$ in~\eqref{eq:x_{t,n}} over the index $t$ can be listed as
\begin{enumerate}
	\item For $t=1,2,\dots,\tau$, collect all values of $\bar{x}_{t,n}[\tilde{n}]$ in~\eqref{eq:x_{t,n}} and compute their corresponding DTFT $\bar{X}_{t,n}(\vartheta)$ in~\eqref{eq:X_{t,n}} for all $n \in \mathcal{M}$. We use them to form $\bar{\bf y}_{t}(\vartheta)$ in~\eqref{eq:y_{t}_bar}.
	\item Compute $\hat{\bf R}_{\bar{y}}(\vartheta)$, for $\vartheta \in [0,1/N)$, using~\eqref{eq:Rybar_hat}.
	\item Based on~\eqref{eq:Rybar_as_rbar_x} and for $\vartheta \in [0,1/N)$, we apply 
	LS reconstruction on $\hat{\bf R}_{\bar{y}}(\vartheta)$ leading to
\vspace{-0.5mm}
		\begin{equation}
				\hat{\bf r}_{\bar{x},LS}(\vartheta)=({\bf R}_c^T{\bf R}_c)^{-1}{\bf R}_c^T\text{vec}(\hat{\bf R}_{\bar{y}}(\vartheta)).
				\label{rxhatbarLS}
				\vspace{-0.5mm}
		\end{equation} 
	\item Based on~\eqref{eq:Rbarx_as_rbar_x} and~\eqref{eq:Rxt_bar}, for $\vartheta \in [0,1/N)$, we compute $\text{vec}(\hat{\bf R}_{\bar{x},LS}(\vartheta))={\bf T}\hat{\bf r}_{\bar{x},LS}(\vartheta)$ and
		\vspace{-0.5mm}
		\begin{equation}
			\hat{\bf R}_{x,LS}(\vartheta)=N^2{\bf B}^{H}\hat{\bf R}_{\bar{x},LS}(\vartheta){\bf B}. 
	  	\label{eq:RxLS}
	  	\vspace{-0.5mm}
	  \end{equation} 	
	  \item Note that the $(i+1)$-th diagonal element of $\hat{\bf R}_{x,LS}(\vartheta)$, i.e., $[\text{diag}(\hat{\bf R}_{x,LS}(\vartheta))]_{i+1}$ is the LS estimate of the $(i+1)$-th diagonal element of ${\bf R}_{x}(\vartheta)$, which according to Remark~1 is given by $E[|X_{t,i}(\vartheta)|^2]$. Based on the definition of AP in Remark~1 and considering~\eqref{eq:Rybar_hat}, we can then formulate the compressive 
	  AP (CAP) of $x_t[\tilde{n}]$ in~\eqref{eq:x_{t,n}} over the index $t$ as
	  \begin{equation}
	  \hat{P}_{x,LS}(\vartheta+\frac{i}{N})=\frac{1}{\tilde{N}}[\text{diag}(\hat{\bf R}_{x,LS}(\vartheta))]_{i+1},
	  \label{CRAP}
	  \end{equation} 
	  for $\vartheta \in [0,1/N)$ and $i=0,1,\dots,N-1$.
\end{enumerate}

Note that, when reconstructing the CAP $\hat{P}_{x,LS}(\vartheta)$ in~\eqref{CRAP}, we introduce additional errors with respect to the AP $\frac{1}{\tilde{N}\tau}\sum_{t=1}^{\tau}|X_t(\vartheta)|^2$ in Remark~1 (including the ones in~\eqref{ExS3P2} and~\eqref{ExS1P1}).
This error emerges during the compression and the LS operation in~\eqref{rxhatbarLS}. 
This issue will be discussed up to some extent in the next section.

\section{Performance Analysis}\label{performance}
\subsection{Bias Analysis}\label{uncorrbins_bias}

The 
bias analysis of the CAP $\hat{P}_{{x},{LS}}(\vartheta)$ in~\eqref{CRAP} with respect to $P_{x}(\vartheta)$ in~\eqref{eq:PowerSpectrum} is given by the following theorem whose proof is available in 
Appendix~\ref{ProofTheorem2}. 
\vspace{0.7mm}
\newline
\hspace*{1mm}{\it Theorem 2: For 
$\vartheta \in [0,1)$, the CAP $\hat{P}_{{x},{LS}}(\vartheta)$ in~\eqref{CRAP} is an asymptotically (with respect to $\tilde{N}$) unbiased estimate of 
$P_{x}(\vartheta)$ in~\eqref{eq:PowerSpectrum}.}
\vspace{-1mm}
\subsection{Variance Analysis}\label{uncorrbins_variance}

We start by recalling that the $(m+1)$-th element of $\bar{\bf y}_{t}(\vartheta)$ in~\eqref{eq:y_{t}_bar} is given by $\bar{X}_{t,n_m}(\vartheta)$. By using~\eqref{eq:X_{t,n}}, we can write the element of $\hat{\bf R}_{\bar{y}}(\vartheta)$ in~\eqref{eq:Rybar_hat} at the $(m+1)$-th row and the $(m'+1)$-th column, for $m,m'=0,1,\dots,M-1$, as
\begin{align}
&[\hat{\bf R}_{\bar{y}}(\vartheta)]_{m+1,m'+1}
=\frac{1}{N^2\tau}\sum_{t=1}^{\tau}\sum_{i=0}^{N-1}\sum_{i'=0}^{N-1}\nonumber\\
&X_{t,i}(\vartheta)X_{t,i'}^*(\vartheta)e^{\frac{j2\pi (n_m i-n_{m'}i')}{N}}.
\label{elementRybarhat}
\end{align}
We continue to evaluate the covariance between the elements of $\hat{\bf R}_{\bar{y}}(\vartheta)$ in~\eqref{elementRybarhat}, which 
is not trivial for a general signal $x_t[\tilde{n}]$ in~\eqref{eq:x_{t,n}}, as it involves the computation of fourth order moments. 
To get a useful insight, let us consider the case when the distribution of 
$x_t[\tilde{n}]$ in~\eqref{eq:x_{t,n}} (and thus also ${X}_{t,i}(\vartheta)$ in~\eqref{elementRybarhat}) is jointly Gaussian. 
In this case, 
the fourth order moment computation is simplified 
by using the results in~\cite{Bar}: If $x_1$, $x_2$, $x_3$, and $x_4$ are jointly (real or complex) 
Gaussian random variables, we have $E[x_1 x_2 x_3 x_4]=E[x_1 x_2]E[x_3 x_4]+E[x_1 x_3]E[x_2 x_4]+E[x_1 x_4]E[x_2 x_3]-2E[x_1]E[x_2]E[x_3]E[x_4]$.
Using this result, 
the covariance between the elements of $\hat{\bf R}_{{\bar{y}}}(\vartheta)$ in~\eqref{elementRybarhat}, 
when $x_t[\tilde{n}]$ in~\eqref{eq:x_{t,n}} is jointly Gaussian, can be shown to be 
\begin{align}
&\text{Cov}[[\hat{\bf R}_{{\bar{y}}}(\vartheta)]_{m+1,m'+1},[\hat{\bf R}_{{\bar{y}}}(\vartheta)]_{a+1,a'+1}]=\frac{1}{N^4\tau^2}\sum_{t=1}^{\tau}\sum_{t'=1}^{\tau}\nonumber \\
&\sum_{i=0}^{N-1}\sum_{i'=0}^{N-1}\sum_{b=0}^{N-1}\sum_{b'=0}^{N-1}e^{\frac{j2\pi (n_m i-n_{m'} i'-n_ab+n_{a'}b')}{N}}\nonumber \\
&\left\{E[{X}_{t,i}(\vartheta){X}_{t',b}^*(\vartheta)]E[{X}_{t,i'}^*(\vartheta){X}_{t',b'}(\vartheta)]
+\right.\nonumber \\
&\left.E[{X}_{t,i}(\vartheta){X}_{t',b'}(\vartheta)]E[{X}_{t,i'}^*(\vartheta){X}_{t',b}^*(\vartheta)]\right\},
\label{eq:CovRycheckbarelementGaussian}
\end{align}
for $\vartheta \in [0,1/N)$ and $m,m',a,a'=0,1,\dots,M-1$, where we also 
assume that $x_t[\tilde{n}]$ in~\eqref{eq:x_{t,n}} has zero mean (see Definition~1). 

Under the above assumptions, we introduce the $M^2 \times M^2$ covariance matrix 
${\boldsymbol \Sigma}_{\hat{R}_{{\bar{y}}}}(\vartheta)=E[\text{vec}(\hat{\bf R}_{{\bar{y}}}(\vartheta))\text{vec}(\hat{\bf R}_{{\bar{y}}}(\vartheta))^H]-E[\text{vec}(\hat{\bf R}_{{\bar{y}}}(\vartheta))]E[\text{vec}(\hat{\bf R}_{{\bar{y}}}(\vartheta))^H]$, whose 
entry at the $(Mm'+m+1)$-th row and the $(Ma'+a+1)$-th column
is given by $\text{Cov}[[\hat{\bf R}_{{\bar{y}}}(\vartheta)]_{m+1,m'+1},[\hat{\bf R}_{{\bar{y}}}(\vartheta)]_{a+1,a'+1}]$ in~\eqref{eq:CovRycheckbarelementGaussian}. 
By recalling that ${\bf R}_{c}$ and ${\bf T}$ are real matrices, we can then compute the $N\times N$ covariance matrix of $\hat{\bf r}_{{\bar{x}},{LS}}(\vartheta)$ in~\eqref{rxhatbarLS} as
\begin{equation}
{\boldsymbol \Sigma}_{\hat{r}_{{\bar{x}},LS}}(\vartheta)
=({\bf R}_{c}^T{\bf R}_{c})^{-1}{\bf R}_{c}^T{\boldsymbol \Sigma}_{\hat{R}_{{\bar{y}}}}(\vartheta){\bf R}_{c}({\bf R}^T_{c}{\bf R}_{c})^{-1},
\label{eq:Covar_scheckhatbarXLS}
\end{equation}
and 
use~\eqref{eq:RxLS} to introduce ${\boldsymbol \Sigma}_{\hat{R}_{{x},LS}}(\vartheta)$ as 
the $N^2\times N^2$ covariance matrix of $\text{vec}(\hat{\bf R}_{{{{x}}},{LS}}(\vartheta))$, which can be written as
\begin{equation}
{\boldsymbol \Sigma}_{\hat{R}_{{x},LS}}(\vartheta)
=N^4({\bf B}^T\otimes{\bf B}^{H}){\bf T}{\boldsymbol \Sigma}_{\hat{r}_{{\bar{x}},LS}}(\vartheta){\bf T}^T({\bf B}^*\otimes{\bf B}),
\label{eq:Covar_ScheckhatXLS}
\end{equation}
for $\vartheta \in [0,1/N)$. Recall from~\eqref{CRAP} that the 
CAP $\hat{P}_{{x},{LS}}(\vartheta+\frac{i}{N})$, for $\vartheta \in [0,1/N)$ and $i=0,1,\dots,N-1$, is given by $\frac{1}{\tilde{N}}[\hat{\bf R}_{{{{x}}},{LS}}(\vartheta)]_{i+1,i+1}$. 
It is then trivial to show that the variance of $\hat{P}_{{x},{LS}}(\vartheta+\frac{i}{N})$ is given by 
\begin{equation}
\text{Var}[\hat{P}_{{x},{LS}}(\vartheta+\frac{i}{N})]=\frac{1}{\tilde{N}^2}[{\boldsymbol \Sigma}_{\hat{R}_{{x},LS}}(\vartheta)]_{Ni+i+1,Ni+i+1},
\label{eq:Var_LS_periodogoram}
\end{equation}
for $\vartheta \in [0,1/N)$ and $i=0,1,\dots,N-1$. 

To get even more insight into this result, we consider a specific case in the next proposition whose proof is provided in Appendix~\ref{ProofPropos1}.
\vspace{1mm}
\newline
\hspace*{1mm}{\it Proposition 1: When $x_t[\tilde{n}]$ in~\eqref{eq:x_{t,n}} contains only circular complex zero-mean Gaussian i.i.d. noise with variance $\sigma^2$,
the covariance between the elements of $\hat{\bf R}_{{\bar{y}}}(\vartheta)$ in~\eqref{elementRybarhat}, for $\vartheta \in [0,1/N)$, is given by
\vspace{-0.5mm}
\begin{align}
&\text{Cov}[[\hat{\bf R}_{\bar{y}}(\vartheta)]_{m+1,m'+1},[\hat{\bf R}_{\bar{y}}(\vartheta)]_{a+1,a'+1}]=\frac{L^2\sigma^4}{\tau}\times\nonumber\\
&\delta[m -{a}]\delta[{m'}-{a'}],\:\:m,m',a,a'=0,1,\dots,M-1.
\label{eq:CovRycheckbarelementGaussian_noise_propos}
\vspace{-0.75mm}
\end{align}}
It is clear from~\eqref{eq:CovRycheckbarelementGaussian_noise_propos} that ${\boldsymbol \Sigma}_{\hat{R}_{{\bar{y}}}}(\vartheta)$ in~\eqref{eq:Covar_scheckhatbarXLS} is then a diagonal matrix and we can 
find from~\eqref{eq:Covar_scheckhatbarXLS}
-\eqref{eq:Var_LS_periodogoram} 
that $\text{Var}[\hat{P}_{{x},{LS}}(\vartheta)] \propto \sigma^4$ or $\text{Var}[\hat{P}_{{x},{LS}}(\vartheta)] \propto {P}^2_{{x}}(\vartheta)$. This observation can be related to a similar result found for the conventional periodogram estimate of white Gaussian noise sampled at Nyquist rate in~\cite{Hayes}.

\vspace{-0.5mm}\subsection{Effect of the Compression Rate on the Variance}
\label{comprate_variance}

In this section, we focus on the impact of the compression rate $M/N$ on the variance analysis by first {\color{blue}defining an $N\times 1$ vector ${\bf w}=[w[0],w[1],\dots,w[N-1]]^T$ containing 
binary entries, with $w[n]=1$ if $n \in \mathcal{M}$ (i.e., the coset with index $n$ is one of the $M$ activated cosets) and $w[n]=0$ if $n \notin \mathcal{M}$. In other words, the entries of ${\bf w}$ indicate which $M$ out of the $N$ cosets are activated.
Let us then focus on~\eqref{rxhatbarLS} 
and} consider the following remark.
\vspace{1mm}
\newline{\color{blue}
\hspace*{1mm}{\it Remark 3: The same argument that leads to Theorem~1 
(see Lemma~1 in~\cite{CAMSAP13}) 
shows that the rows of ${\bf R}_{c}$ are given by the $((g-f)\text{ mod }N+1)$-th rows of ${\bf I}_N$, for all $f,g \in \mathcal{M}$. As a result, ${\bf R}_{c}^T{\bf R}_{c}$ 
is an $N\times N$ diagonal matrix. Denote the value of the $\kappa$-th diagonal element of ${\bf R}_{c}^T{\bf R}_{c}$ as ${\gamma_\kappa}$. We can then show that ${\gamma_\kappa}$ is given by
\begin{equation}
\gamma_\kappa=\sum_{n=0}^{N-1}w[(n+\kappa-1)\text{ mod }N]w[n],\:\:\kappa=1,2,\dots,N.
\label{eq:gamma_kappa}
\end{equation}
The proof of~\eqref{eq:gamma_kappa} is available in Appendix~\ref{ProofOfGammakappa}. 
Using~\eqref{eq:gamma_kappa}, we can also show that $\gamma_\kappa$ gives the number of times the $\kappa$-th row of ${\bf I}_N$ appears in ${\bf R}_{c}$, i.e., the number of pairs $(g,f)$ that lead to $(g-f)\text{ mod }N+1=\kappa$. As we have $|\mathcal{M}|=M$, we can find that 
$\sum_{\kappa=1}^N\gamma_\kappa=M^2$ and $\gamma_1=M$.}} 
\vspace{0.5mm}
\newline
Using Remark~3, we then 
formulate the following theorem whose proof is available in Appendix~\ref{ProofTheorem3}. 
\newline
\hspace*{1mm}{\it Theorem 3: When $x_t[\tilde{n}]$ in~\eqref{eq:x_{t,n}} contains only circular complex zero-mean Gaussian i.i.d. noise with variance $\sigma^2$, the variance of the 
CAP $\hat{P}_{{x},{LS}}(\vartheta+\frac{i}{N})$ in~\eqref{eq:Var_LS_periodogoram}, for $\vartheta \in [0,1/N)$ and $i=0,1,\dots,N-1$, is given by
\begin{equation}
\text{Var}[\hat{P}_{{x},{LS}}(\vartheta+\frac{i}{N})]=\frac{\sigma^4}{M\tau}+\frac{\sigma^4}{\tau}\sum_{n=1}^{N-1}\frac{1}{\gamma_{n+1}}.
\label{eq:VarPxLSwhitenoisetheo}
\end{equation}}
Note how~\eqref{eq:VarPxLSwhitenoisetheo} 
relates 
$M$ and $N$ to $\text{Var}[\hat{P}_{{x},{LS}}(\vartheta)]$ for circular complex zero-mean Gaussian i.i.d. noise and $\vartheta \in [0,1)$. 
Recalling from Remark~3 that $\sum_{n=1}^{N-1}\gamma_{n+1}=M^2-M$, 
we can find that, for a given $N$, 
a stronger compression rate (smaller $M/N$) 
tends to lead to a larger $\text{Var}[\hat{P}_{{x},{LS}}(\vartheta)]$. 
{\color{blue}Based on~\eqref{eq:gamma_kappa} and~\eqref{eq:VarPxLSwhitenoisetheo},
it is of interest to find the binary values of $\{w[n]\}_{n=0}^{N-1}$ 
(or equivalently the cosets $n_m\in\mathcal{M}$) that minimize $\text{Var}[\hat{P}_{{x},{LS}}(\vartheta)]$ for a given $M$. This will generally lead to a non-convex optimization problem, which is difficult to solve, although it is clear that the solution will force the values of $\{\gamma_{n+1}\}_{n=1}^{N-1}$ to be as equal as possible. 
Alternatively, we can also put a constraint on $\text{Var}[\hat{P}_{{x},{LS}}(\vartheta+\frac{i}{N})]$ in~\eqref{eq:VarPxLSwhitenoisetheo} and find the binary values $\{w[n]\}_{n=0}^{N-1}$ that minimize the compression rate $M/N$. This however, will again lead to a non-convex optimization problem that is difficult to solve.
Note that, although finding ${\bf w}$ that minimizes $M/N$ for a given $\text{Var}[\hat{P}_{{x},{LS}}(\vartheta)]$ in~\eqref{eq:VarPxLSwhitenoisetheo} or the one that minimizes $\text{Var}[\hat{P}_{{x},{LS}}(\vartheta)]$ for a given $M/N$ is not trivial, 
the solution will always have to satisfy the identifiability condition in Theorem~1. This is because we can show that if the identifiability condition is not satisfied, some 
$\gamma_{n}$ in~\eqref{eq:VarPxLSwhitenoisetheo} will be zero and thus $\text{Var}[\hat{P}_{{x},{LS}}(\vartheta)]$ in~\eqref{eq:VarPxLSwhitenoisetheo} will have an infinite value.} 

The analysis of the effect of $M/N$ on $\text{Var}[\hat{P}_{{x},{LS}}(\vartheta)]$ for a general Gaussian signal $x_t[\tilde{n}]$, however, is difficult since it is clear from~\eqref{eq:CovRycheckbarelementGaussian} 
that $\text{Var}[\hat{P}_{{x},{LS}}(\vartheta)]$ for this case depends on the unknown statistics of $x_t[\tilde{n}]$. This is also true for a more general signal. 

\vspace{-0.5mm}\subsection{Asymptotic Performance Analysis}\label{asymp_variance}

We now discuss the asymptotic behaviour of the performance of the 
CAP $\hat{P}_{{x},{LS}}(\vartheta)$. We start by noting that 
Assumption~1 ensures that $\hat{\bf R}_{{\bar{y}}}(\vartheta)$ in~\eqref{eq:Rybar_hat} is a consistent estimate of ${\bf R}_{{\bar{y}}}(\vartheta)$ in~\eqref{eq:Rybar_as_rbar_x} i.e., $\hat{\bf R}_{{\bar{y}}}(\vartheta)$ converges to ${\bf R}_{{\bar{y}}}(\vartheta)$ 
as $\tau$ approaches $\infty$. As it is clear from~\eqref{rxhatbarLS} and~\eqref{eq:RxLS} that $\hat{\bf R}_{x,LS}(\vartheta)$ is linearly related to $\hat{\bf R}_{{\bar{y}}}(\vartheta)$, it is easy to show that $\hat{\bf R}_{x,LS}(\vartheta)$ converges to ${\bf R}_{x}(\vartheta)$ in~\eqref{eq:Rxt_bar} 
as $\tau$ approaches $\infty$. This implies that the CAP $\hat{P}_{x,LS}(\vartheta+\frac{i}{N})$ in~\eqref{CRAP} also converges to $\frac{1}{\tilde{N}}[\text{diag}({\bf R}_{x}(\vartheta))]_{i+1}=\frac{1}{\tilde{N}}E[|X_t(\vartheta+\frac{i}{N})|^2]$, for $\vartheta \in [0,1/N)$ and $i=0,1,\dots,N-1$, 
as $\tau$ approaches $\infty$. Since $x_t[\tilde{n}]$ in~\eqref{eq:x_{t,n}} is an observation of the true process $x[\tilde{n}]$ in~\eqref{eq:PowerSpectrum}, $\hat{P}_{x,LS}(\vartheta)$ will converge to ${P}_{x}(\vartheta)$ in~\eqref{eq:PowerSpectrum} 
if both $\tau$ and $\tilde{N}$ (or $L$ for a fixed $N$) approach $\infty$. 

{\color{blue}\subsection{Complexity Analysis}\label{complexity}

Let us now compare the complexity of our CAP approach with an existing state-of-the-art approach to tackle similar problems. 
We compare our CAP approach with a method that 
reconstructs 
$X_t(\vartheta)$ (instead of the periodogram), for $\vartheta \in [0,1)$ and all $t=1,2,\dots,\tau$, from compressive measurements. The reconstruction of $\{X_t(\vartheta)\}_{t=1}^{\tau}$, for $\vartheta \in [0,1)$, is performed by reconstructing $\{{\bf x}_{t}(\vartheta)\}_{t=1}^{\tau}$ in~\eqref{eq:x_{t}_bar} from $\{\bar{\bf y}_{t}(\vartheta)\}_{t=1}^{\tau}$ in~\eqref{eq:y_{t}_bar}, for $\vartheta \in [0,1/N)$, using the Regularized M-FOCUSS (RM-FOCUSS) approach of~\cite{BhaskarRao}.
We then use the reconstructed $\{{\bf x}_{t}(\vartheta)\}_{t=1}^{\tau}$, for $\vartheta \in [0,1/N)$, either to compute the periodogram or to compute the energy at 
$\vartheta \in [0,1)$ and to detect the existence of 
active user signals.
Note that RM-FOCUSS is designed to treat $\{\bar{\bf y}_{t}(\vartheta)\}_{t=1}^{\tau}$, for each 
$\vartheta$, as multiple measurement vectors (MMVs) and exploit the assumed joint sparsity structure in $\{{\bf x}_{t}(\vartheta)\}_{t=1}^{\tau}$.}

{\color{blue}Table~\ref{tab:compute_complex} summarizes the computational complexity of 
CAP and 
RM-FOCUSS 
(see~\cite{BhaskarRao} for more details). 
Note that Table~\ref{tab:compute_complex} only describes the computational complexity of RM-FOCUSS for a single iteration. The number of RM-FOCUSS iterations depends on the 
convergence criterion parameter (labeled as $\delta$ in~\cite{BhaskarRao}). Hence, we can argue that our CAP approach is simpler than RM-FOCUSS. Moreover, in RM-FOCUSS, we also need to determine a proper 
regularization parameter (labeled as $\lambda$ in~\cite{BhaskarRao}), which is generally not a trivial task. Note that we also compare the detection performance of the two methods in the sixth experiment of Section~\ref{simulation_uncorrbins}. Note that 
the reconstruction of ${\bf x}_{t}(\vartheta)$ from $\bar{\bf y}_{t}(\vartheta)$ is also considered in~\cite{Eldar} but it 
only considers the single-sensor case. 
\begin{table}[ht]
	\caption{
	Computational complexity of the CAP approach and the RM-FOCUSS of~\cite{BhaskarRao} for a given frequency point $\vartheta \in [0,1/N)$.}
	\centering
	\vspace{-1mm}
		\begin{tabular}{| c | c |}
			\hline
			\multicolumn{2}{|c|}{CAP approach}\\
			\hline
    Computation steps & Computational complexity\\ \hline
    Computation of $\hat{\bf R}_{\bar{y}}(\vartheta)$ in~\eqref{eq:Rybar_hat}& $\mathcal{O}(M^2\tau)$\\\hline   
    Computation of ${\bf R}_c^T{\bf R}_c$ in~\eqref{rxhatbarLS} & $\mathcal{O}(N^2M^2)$\\\hline 
    Inversion of ${\bf R}_c^T{\bf R}_c$ in~\eqref{rxhatbarLS} & $\mathcal{O}(N^3)$\\\hline
    Multiplication between $({\bf R}_c^T{\bf R}_c)^{-1}$ & $\mathcal{O}(N^2M^2)+$\\  
    and ${\bf R}_c^T\text{vec}(\hat{\bf R}_{\bar{y}}(\vartheta))$ in~\eqref{rxhatbarLS} &$\mathcal{O}(NM^2)$\\\hline
    Computation of~\eqref{eq:RxLS} (recall that & $\mathcal{O}(N\text{ log }N)$\\
    ${\bf B}$ in~\eqref{eq:RxLS} is an IDFT matrix) & \\\hline
		Total & $\mathcal{O}(N^3)+\mathcal{O}(N^2M^2)$\\
		& $+\mathcal{O}(M^2\tau)$ \\\hline
   \multicolumn{2}{|c|}{RM-FOCUSS of~\cite{BhaskarRao} (per iteration)}\\\hline
    Computation steps & Computational complexity\\ \hline
    Computation of $\ell_2$-norm of & $\mathcal{O}(N\tau)$\\
    each row of an $N\times\tau$ matrix & \\\hline
    Multiplication between an $M\times N$&$\mathcal{O}(N^2M)$\\
    matrix and an $N\times N$ matrix & \\\hline
    Multiplication between an $M\times N$ &$\mathcal{O}(NM^2)$\\
    matrix and an $N\times M$ matrix & \\\hline
    Inversion of an $M\times M$ matrix &$\mathcal{O}(M^3)$\\\hline
    Multiplication between an $N\times M$ &$\mathcal{O}(NM^2)$\\
    matrix and an $M\times M$ matrix & \\\hline
    Multiplication between an $N\times M$ &$\mathcal{O}(NM\tau)$\\
    matrix and an $M\times \tau$ matrix &\\\hline
    Multiplication between an $N\times N$ &$\mathcal{O}(N^2\tau)$\\
    matrix and an $N\times \tau$ matrix  &\\\hline
    Total & $\mathcal{O}(N^2M)+\mathcal{O}(M^3)+$\\
    & $\mathcal{O}(N^2\tau)+\mathcal{O}(NM^2)$\\ 
    & $+\mathcal{O}(NM\tau)$\\\hline   
    \end{tabular}
\label{tab:compute_complex}
\end{table}} 

\section{Multi-cluster Scenario}\label{CaseC2}

Recall that the ergodicity assumption on ${\bf x}_t(\vartheta)$ 
in Assumption~1 
requires the statistics of ${\bf x}_t(\vartheta)$ to be the same along index $t$. 
Let us now consider the case where we have $D$ clusters of $\tau$ time indices in P1 or of $\tau$ sensors in P2 such that ${\bf x}_t(\vartheta)$ is ergodic and its statistics do not change 
only along index $t$ within a cluster. 
We can then consider 
Assumption~1 
and the resulting case considered in Sections~\ref{uncorr_bins_system_model}-\ref{performance} as a special case of this multi-cluster scenario with $D=1$. We introduce the correlation matrix of ${\bf x}_t(\vartheta)$ and $\bar{\bf y}_t(\vartheta)$ for all indices $t$ belonging to cluster $d$ as ${\bf R}_{x,d}(\vartheta)$ and ${\bf R}_{\bar{y},d}(\vartheta)$, respectively, with $d=0,1,\dots,D-1$. We can then repeat all the steps of Sections~\ref{uncorr_bins_system_model}-\ref{performance}
for each cluster. More precisely, we can follow~\eqref{eq:Rybar_hat} and define the estimate of ${\bf R}_{\bar{y},d}(\vartheta)$ as $\hat{\bf R}_{\bar{y},d}(\vartheta)$, which is computed by averaging the outer-product of $\bar{\bf y}_t(\vartheta)$ over indices $t$ belonging to cluster $d$.
Then, we apply~\eqref{rxhatbarLS}-\eqref{CRAP} on $\hat{\bf R}_{\bar{y},d}(\vartheta)$ to obtain $\hat{\bf R}_{x,LS,d}(\vartheta)$ and the CAP for cluster $d$, i.e., $\hat{P}_{x,LS,d}(\vartheta)$. Also note that the bias and variance analysis in Section~\ref{performance} is also valid for each cluster in this section.

We might then be interested in the averaged statistics over the clusters, i.e., $\frac{1}{D}\sum_{d=0}^{D-1}{\bf R}_{x,d}(\vartheta)$. Since $\frac{1}{D}\sum_{d=0}^{D-1}\hat{\bf R}_{\bar{y},d}(\vartheta)$ is a consistent estimate of $\frac{1}{D}\sum_{d=0}^{D-1}{\bf R}_{\bar{y},d}(\vartheta)$, we can then consider the resulting $\frac{1}{D}\sum_{d=0}^{D-1}\hat{\bf R}_{{x},{LS},d}(\vartheta)$ as a valid LS estimate of $\frac{1}{D}\sum_{d=0}^{D-1}{\bf R}_{x,d}(\vartheta)$. Defining the theoretical spectral representation of the power at cluster $d$ as ${P}_{x,d}(\vartheta)$, we can then apply Theorem~2 for each cluster to conclude that $\frac{1}{D}\sum_{d=0}^{D-1}\hat{P}_{{x},{LS},d}(\vartheta)$ is an asymptotically (with respect to $\tilde{N}$) unbiased estimate of $\frac{1}{D}\sum_{d=0}^{D-1} P_{x,d}(\vartheta)$. This multi-cluster scenario is of interest for P2 when we have clusters of wireless sensors sensing user signals where the signal from each user experiences the same fading statistics (the same path loss and shadowing) on its way towards the sensors belonging to the same cluster. However, the fading statistics experienced by the signal between the user location and different clusters are not the same. For P1, the multi-cluster scenario implies that the array sensing time can be grouped into 
multiple clusters of time indices where the 
signal statistics do not vary 
along the time within the cluster but 
they vary across different clusters.

\vspace{-1mm}
\section{
Correlated Bins}\label{correlatedbins}

When the bin size is reduced by increasing $N$ in~\eqref{eq:X_{t,n}}, the received spectra at two frequencies or angles, which are separated by more 
than the size of the bin, might still be correlated. 
In this case, ${\bf R}_{x}(\vartheta)$ and ${\bf R}_{\bar{x}}(\vartheta)$ in~\eqref{eq:Rxt_bar} are respectively not a diagonal and circulant matrix anymore, and 
the temporal and spatial compression of Section~\ref{uncorr_bins_compression} cannot be performed without violating the identifiability of ${\bf r}_{\bar{x}}(\vartheta)$ in~\eqref{eq:Rybar_as_rbar_x}. 
This section proposes a solution when this situation occurs under 
Assumption~1 and the single-cluster scenario (it does not apply to the multi-cluser scenario of Section~\ref{CaseC2}).
Let us organize $\tau$ indices $t$ into several groups 
and write $t$ as $t=pZ+z+1$ with $p=0,1,\dots,P-1$ and $z=0,1,\dots,Z-1$, where $Z$ and $P$ represent the total number of groups 
and the number of indices belonging to a group, respectively. 
Writing $\bar{\bf y}_t(\vartheta)$ and $\bar{\bf x}_t(\vartheta)$ at $t=pZ+z+1$ as $\bar{\bf y}_{p,z}(\vartheta)$ and $\bar{\bf x}_{p,z}(\vartheta)$, we can
introduce for each $z$ a compression similar to~\eqref{eq:y_{t}_bar} as
\vspace{-1mm}
\begin{equation}
\bar{\bf y}_{p,z}(\vartheta)={\bf C}_{z}\bar{\bf x}_{p,z}(\vartheta),\quad \vartheta \in [0,1/N),
\label{eq:y_pz_bar}
\vspace{-1mm}
\end{equation}
where ${\bf C}_{z}$ is the $M\times N$ selection matrix for the $z$-th group of indices whose rows are also selected from the rows of ${\bf I}_N$. 
Next, 
we compute the correlation matrix of $\bar{\bf y}_{p,z}(\vartheta)$ in~\eqref{eq:y_pz_bar}, i.e., ${\bf R}_{\bar{y}_{z}}(\vartheta)=E[\bar{\bf y}_{p,z}(\vartheta)\bar{\bf y}^H_{p,z}(\vartheta)]$, for $z=0,1,\dots,Z-1$, as 
\vspace{-2mm}
\begin{equation}
{\bf R}_{\bar{y}_{z}}(\vartheta)
={\bf C}_{z}E[\bar{\bf x}_{p,z}(\vartheta)\bar{\bf x}^H_{p,z}(\vartheta)]{\bf C}^T_{z}={\bf C}_{z}{\bf R}_{\bar{x}}(\vartheta){\bf C}^T_{z},
\label{eq:Ry_z_bar}
\vspace{-1mm}
\end{equation}
with ${\bf R}_{\bar{x}}(\vartheta)=E[\bar{\bf x}_{p,z}(\vartheta)\bar{\bf x}^H_{p,z}(\vartheta)]$, for all $p,z$, 
as Assumpti-on~1 requires that the statistics of $\bar{\bf x}_{t}(\vartheta)$ do not vary with $t$.

\begin{figure}[t]
				\centering
        \includegraphics[width=0.48\textwidth]{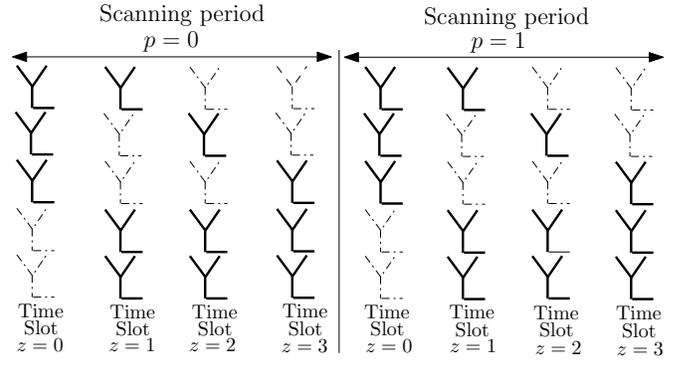}\vspace{-2mm}
        \caption{The 
        DLA model used in problem P1 when the bins are correlated with $M=3$, $N=5$, $P=2$, and $Z=4$. Solid lines and dashed-dotted lines indicate active and inactive antennas, respectively.}        
        \label{fig:SystemModelCorrP1}
\end{figure}

\begin{figure}[t]
				\centering
        \includegraphics[width=0.48\textwidth]{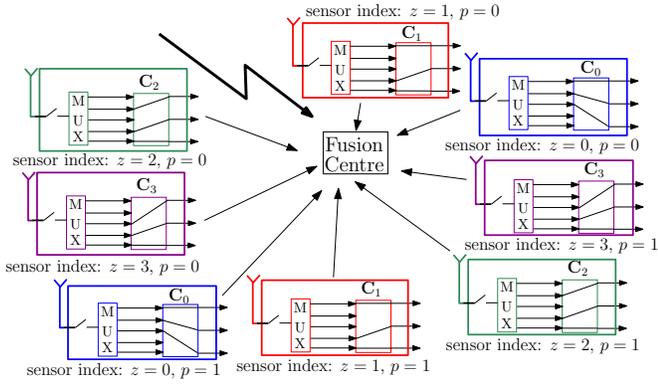}\vspace{-0.5mm}
        \caption{The model for problem P2 when the bins are correlated with $M=3$, $N=5$, $P=2$, and $Z=4$. For simplicity, we illustrate the multi-coset sampling as a Nyquist-rate sampling followed by a multiplexer and a switch that performs sample selection based on ${\bf C}_z$. Sensors in the same group have the same colour. For example, sensors in group $z=0$ collect the samples at the cosets with coset indices $0$,$1$, and $2$.}        
        \label{fig:SystemModelCorrP2}\vspace{-5mm}
\end{figure}

Let us interpret the above model 
for problems P1 and P2. For P1,~\eqref{eq:y_pz_bar} implies that we split the 
array scanning time $\tau$ into $P$ scanning periods, each of which consists of $Z$ time slots. 
It is 
clear from~\eqref{eq:y_pz_bar} that, in different time slots per scanning period, different sets of $M$ ULSs out of $N$ available ULSs in the underlying ULA are activated leading to a dynamic linear array (DLA). This DLA model has actually been introduced in~\cite{ElsevierDOA} though it is originally designed to estimate the DOA of more sources than active antennas, where the sources can be highly correlated.
Here, the indices of the selected rows of ${\bf I}_N$ used to form ${\bf C}_z$ correspond to the indices of the active ULSs at time slot $z$, 
the set of $M$ active ULSs in a given time slot $z$ is the same across different scanning periods, 
and 
the number of received time samples per antenna in a time slot is one. Fig.~\ref{fig:SystemModelCorrP1} shows an example of this DLA 
model. 
For P2,~\eqref{eq:y_pz_bar} implies that  
$\tau$ 
sensors are organized into $Z$ groups of $P$ sensors, where the same 
sampling pattern is adopted by all sensors within the same group and where different groups employ different sampling patterns. 
The indices of the active cosets used by group $z$ then correspond to the indices of the selected rows of ${\bf I}_N$ used to construct ${\bf C}_z$. Fig.~\ref{fig:SystemModelCorrP2} shows an example of the model for problem P2.

Since it turns out that the mathematical model in~\cite{ElsevierDOA} is applicable for both P1 and P2, we can then follow~\cite{ElsevierDOA}, rewrite~\eqref{eq:Ry_z_bar} for $z=0,1,\dots,Z-1$ as 
\vspace{-0.75mm}
\begin{equation*}
{\bf r}_{\bar{y}_z}(\vartheta)=\text{vec}({\bf R}_{\bar{y}_z}(\vartheta))=({\bf C}_{z}\otimes{\bf C}_{z})\text{vec}({\bf R}_{\bar{x}}(\vartheta)),
\label{eq:vec_Ry_z_bar}
\vspace{-0.72mm}
\end{equation*}
combine ${\bf r}_{\bar{y}_z}(\vartheta)$ for all $z$ 
into ${\bf r}_{\bar{y}}(\vartheta)=[{\bf r}^T_{\bar{y}_0}(\vartheta),{\bf r}^T_{\bar{y}_1}(\vartheta),\dots,$ ${\bf r}^T_{\bar{y}_{Z-1}}(\vartheta)]^T$, and write ${\bf r}_{\bar{y}}(\vartheta)$ as
\vspace{-0.5mm}
\begin{equation}
{\bf r}_{\bar{y}}(\vartheta)={\boldsymbol \Psi}\text{vec}({\bf R}_{\bar{x}}(\vartheta)),
\label{eq:ry_bar_vartheta}
\vspace{-0.5mm}
\end{equation}
with ${\boldsymbol \Psi}$ an $M^2Z\times N^2$ matrix given by 
\vspace{-0.5mm}
\begin{equation}
{\boldsymbol \Psi}=[({\bf C}_{0}\otimes{\bf C}_{0})^T,
\dots,({\bf C}_{Z-1}\otimes{\bf C}_{Z-1})^T]^T.
\label{eq:Psi}
\vspace{-0.5mm}
\end{equation}
We can solve for $\text{vec}({\bf R}_{\bar{x}}(\vartheta))$ from ${\bf r}_{\bar{y}}(\vartheta)$ in~\eqref{eq:ry_bar_vartheta} using LS if ${\boldsymbol \Psi}$ in~\eqref{eq:Psi} has full column rank. 
It has been shown in~\cite{ElsevierDOA} that ${\boldsymbol \Psi}$ has full column rank if and only if {\it each possible pair of two different rows} of ${\bf I}_N$ is simultaneously used in {\it at least one} of the matrices 
$\{{\bf C}_z\}_{z=0}^{Z-1}$. 
In P1, this implies that each possible combination of two ULSs in the underlying ULA should be active in at least one time slot per scanning period. 
In P2, this implies that each possible pair of two cosets (out of $N$ possible cosets) should be simultaneously used by at least one group of sensors.
Observe how the 
DLA model in Fig.~\ref{fig:SystemModelCorrP1} and the model in Fig.~\ref{fig:SystemModelCorrP2} satisfy this requirement.
Once $\text{vec}({\bf R}_{\bar{x}}(\vartheta))$ is reconstructed, we follow the procedure in Section~\ref{uncorrbinsreconstruct} to reconstruct ${\bf R}_{{x}}(\vartheta)
=E[{\bf x}_{p,z}(\vartheta){\bf x}^H_{p,z}(\vartheta)]$ from ${\bf R}_{\bar{x}}(\vartheta)$.

In practice, to approximate the expectation operation in computing ${\bf R}_{\bar{y}_z}(\vartheta)$ in~\eqref{eq:Ry_z_bar}, 
we propose to take an average over $\bar{\bf y}_{p,z}(\vartheta)$ at different scanning periods $p$ for P1 or at $P$ sensors in 
group $z$ for P2, i.e., $\hat{\bf R}_{\bar{y}_z}(\vartheta)=\frac{1}{P}\sum_{p=0}^{P-1}\bar{\bf y}_{p,z}(\vartheta)\bar{\bf y}^H_{p,z}(\vartheta)$. Introducing $\hat{\bf r}_{\bar{y}_z}(\vartheta)=\text{vec}(\hat{\bf R}_{\bar{y}_z}(\vartheta))$, the LS reconstruction is then applied to $\hat{\bf r}_{\bar{y}}(\vartheta)=[\hat{\bf r}^T_{\bar{y}_0}(\vartheta),\hat{\bf r}^T_{\bar{y}_1}(\vartheta),\dots,\hat{\bf r}^T_{\bar{y}_{Z-1}}(\vartheta)]^T$.

\section{Numerical Study}\label{numerical}
\subsection{Uncorrelated Bins}\label{simulation_uncorrbins}

In this section, we simulate 
the estimation and detection performance of the 
CAP approach for the uncorrelated bins case discussed in Sections~\ref{uncorr_bins_system_model}-\ref{CaseC2}. 
To keep the study general, in this section, we generally simulate the multi-cluster scenario of Section~\ref{CaseC2}. 
In our first experiment, we consider problem P2 and have $\tilde{N}=3060$, $L=170$, and $N=18$. Each sensor collects $M=5$ samples out of every $N=18$ possible samples based on a periodic length-$17$ minimal circular sparse ruler with $\mathcal{M}=\{0,1,4,7,9\}$. 
This is identical to forming a $5 \times 18$ matrix ${\bf C}$ in~\eqref{eq:y_{t}_bar} 
by selecting 
the rows of ${\bf I}_{18}$ based on $\mathcal{M}$. 
The resulting ${\bf R}_c$ in~\eqref{eq:Rybar_as_rbar_x} has full column rank and we have a compression rate of $M/N=0.28$. We consider $K=6$ user signals whose frequency bands are given in Table~\ref{tab:experiment1} together with the power at each band normalized by frequency. We generate these signals by passing six circular complex zero-mean Gaussian i.i.d. noise signals through different digital filters having $200$ taps where the location of the unit-gain passband of the filter for each signal corresponds to the six different active bands. We set the variances of these noise signals based on the desired user signal powers in Table~\ref{tab:experiment1}. We assume $D=2$ clusters of $\tau=100$ unsynchronized sensors, which means that, at a given point in time, different sensors observe different parts of the user signals. 
To simplify the experiment, the correlation between the different parts of the user signals observed by different sensors is assumed to be negligible such that they can be viewed as independent realizations of the user signals. 
The spatially and temporally white noise 
has a variance of $\sigma^2=7$ dBm. The signal of each user received by different sensors is assumed to pass through different and uncorrelated fading channels $H_t^{(k)}(\vartheta)$. Note however that the signal from a user received by sensors within 
the same cluster is assumed to suffer from the same path loss and shadowing. The amount of path loss experienced between each user and each cluster listed in Table~\ref{tab:experiment1} 
includes the shadowing to simplify the simulation. We simulate small-scale Rayleigh fading on top of the path loss 
by generating the channel frequency response based on a zero-mean complex Gaussian distribution with variance given by the 
path loss in Table~\ref{tab:experiment1}. We assume flat fading in each band. 

\begin{figure}[t]
				\centering
        \includegraphics[width=0.49\textwidth]{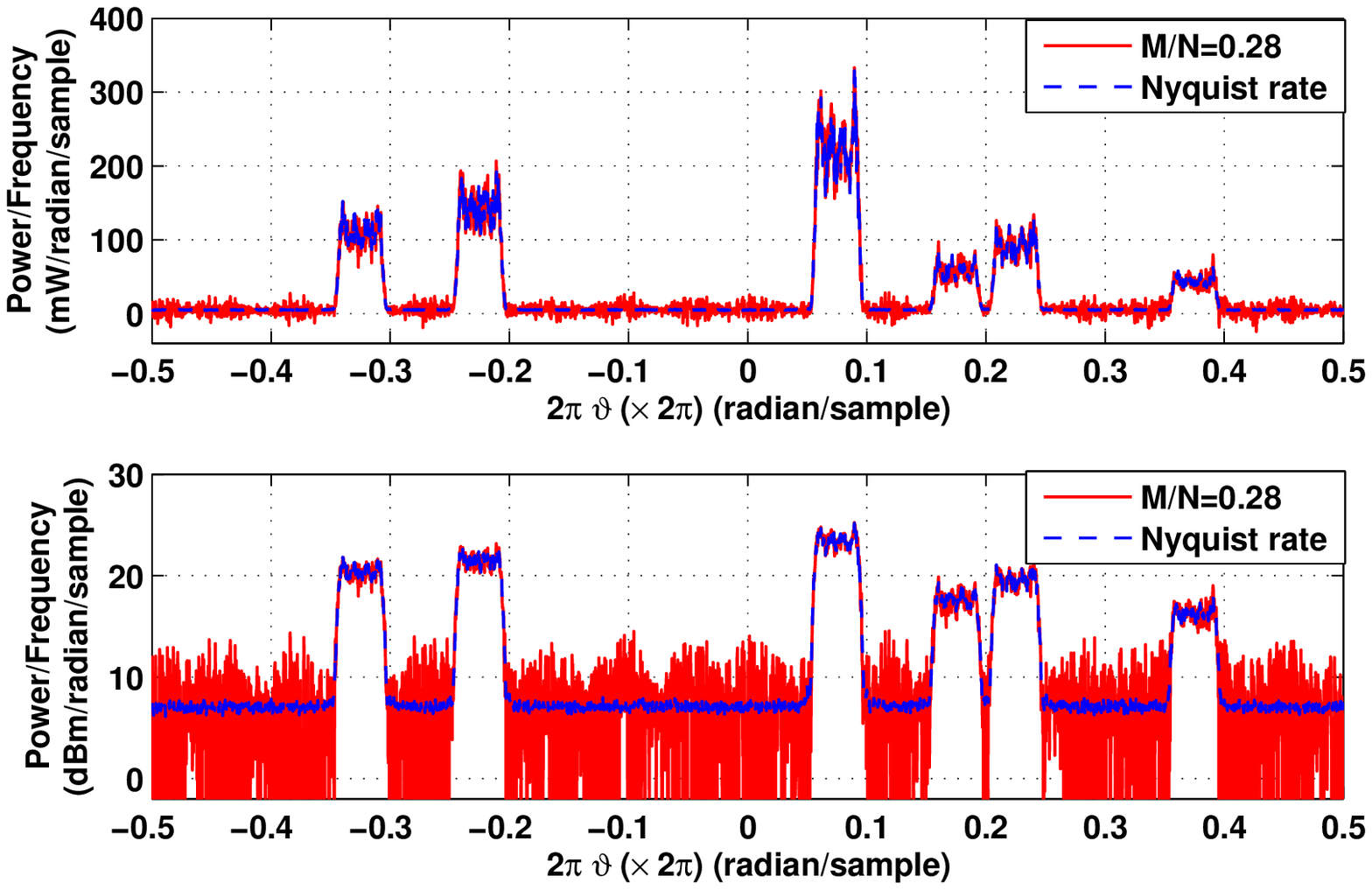}\vspace{-2.5mm}
        \caption{The 
        CAP and the NAP of the faded user signals for the first experiment (unsynchronized sensors) as a function of frequency in a linear scale (top) and logarithmic scale (bottom).}        
        \label{fig:DisplayExperiment1}
				\centering
        \includegraphics[width=0.50\textwidth]{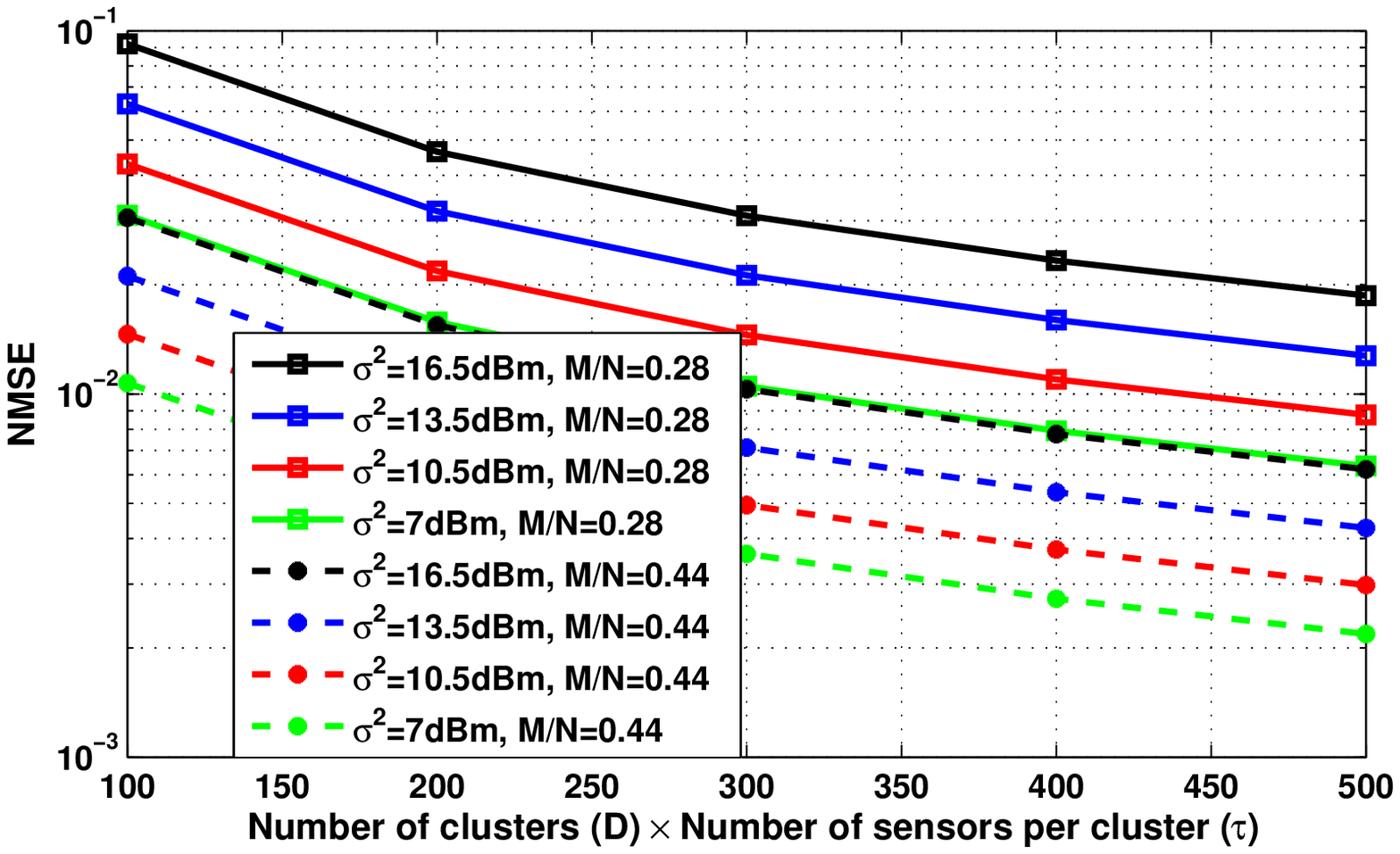}\vspace{-2mm}
        \caption{The NMSE between the 
        CAP and the 
        NAP for the first experiment (unsynchronized sensors).}
        \label{fig:NMSEExperiment1}\vspace{-5mm}
\end{figure}

Fig.~\ref{fig:DisplayExperiment1} shows the CAP of the faded user signals received at the 
sensors. As a benchmark, we provide the Nyquist-rate based AP (NAP), 
which is obtained when all sensors collect all the $\tilde{N}$ samples. 
With respect to the 
NAP, the degradation in the quality of the 
CAP is acceptable despite a strong compression, although more leakage is introduced in the unoccupied band. Next, we perform 1000 Monte Carlo 
runs and vary the number of sensors per cluster $\tau$, the noise variance at each sensor $\sigma^2$, and 
$M/N$ (see Fig.~\ref{fig:NMSEExperiment1}). 
In Fig.~\ref{fig:NMSEExperiment1}, the compression rate of $M/N=0.44$ is implemented by 
activating three extra cosets, i.e., $\{2,12,14\}$ (which we picked randomly). 
Fig.~\ref{fig:NMSEExperiment1} shows the normalized mean square error (NMSE) of the 
CAP with respect to the 
NAP and indicates that increasing $M/N$ 
by a factor of less than two significantly improves the estimation quality. Having more sensors $\tau$ also improves the estimation quality. 
Also observe that the compression introduces a larger NMSE for a larger noise power. 
\begin{table}[t]
	\caption{The frequency band and the power of the users signal and the experienced path loss in the first, second, and third experiments.}
	\centering
	\vspace{-1mm}
		\begin{tabular}{| c | c | c | c |}
			\hline
    User band & Power/freq. &Path loss at&Path loss at\\ 
		 (rad/sample) &(per rad/sample)&cluster 1&cluster 2\\ \hline
    $[-0.69\pi,-0.61\pi]$ & $38$ dBm& $-17$ dB& $-19$ dB\\ \hline
    $[-0.49\pi,-0.41\pi]$ & $40$ dBm& $-20$ dB& $-18$ dB\\ \hline
    $[0.11\pi,0.19\pi]$ & $34$ dBm& $-12$ dB& $-10$ dB\\ \hline
    $[0.31\pi,0.39\pi]$ & $34$ dBm& $-16$ dB& $-18$ dB\\ \hline
    $[0.41\pi,0.49\pi]$ & $32$ dBm& $-14$ dB& $-12$ dB\\ \hline
    $[0.71\pi,0.79\pi]$ & $35$ dBm& $-18$ dB& $-20$ dB\\ \hline
    \end{tabular}
\label{tab:experiment1}
\end{table}

We can also re-interpret the first experiment 
for problem P1. 
In P1, the first experiment implies that $M=5$ ULSs (whose indices are indicated by $\mathcal{M}$) out of $N=18$ ULSs are activated leading to a periodic circular MRA. Table~\ref{tab:experiment1} then gives the angular bands of the $K=6$ user signals 
and the power for each band normalized by the angle. For P1, the first experiment also implies that each user transmits temporally independent signals and that the signals from different users $k$ pass through statistically different and uncorrelated time-varying fading channels $H_t^{(k)}(\vartheta)$ on their way towards the receiving array. For each user $k$, the fading statistics 
remain constant within each cluster of time indices but the fading realization is temporally independent. 

\begin{figure}[t]
				\centering
        \includegraphics[width=0.49\textwidth]{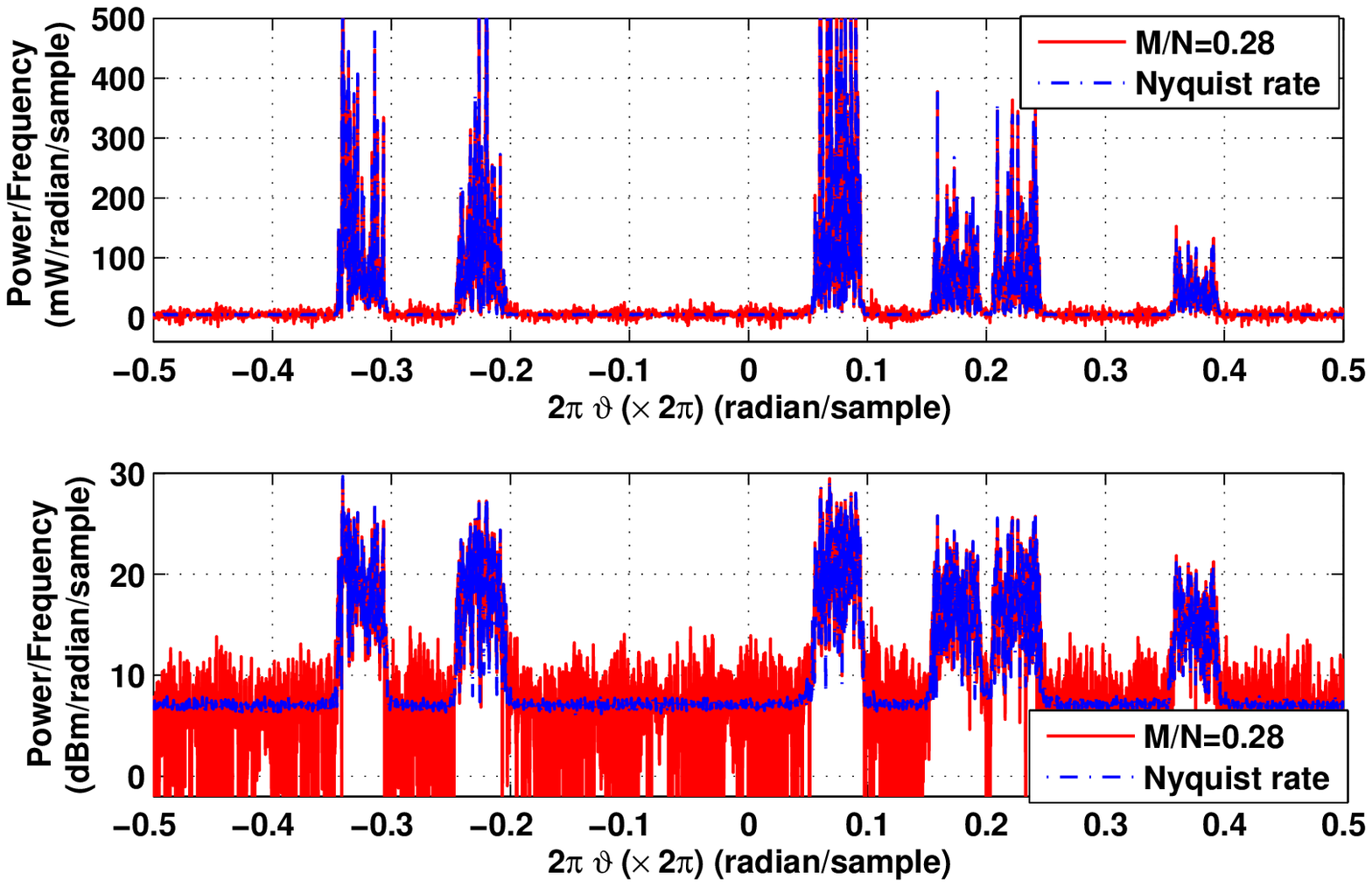}\vspace{-2mm}
        \caption{The 
        CAP and the NAP of the faded user signals for the second experiment (synchronized sensors) as a function of frequency in a linear scale (top) and logarithmic scale (bottom).}
        \label{fig:DisplayExperiment2}
				\centering
        \includegraphics[width=0.49\textwidth]{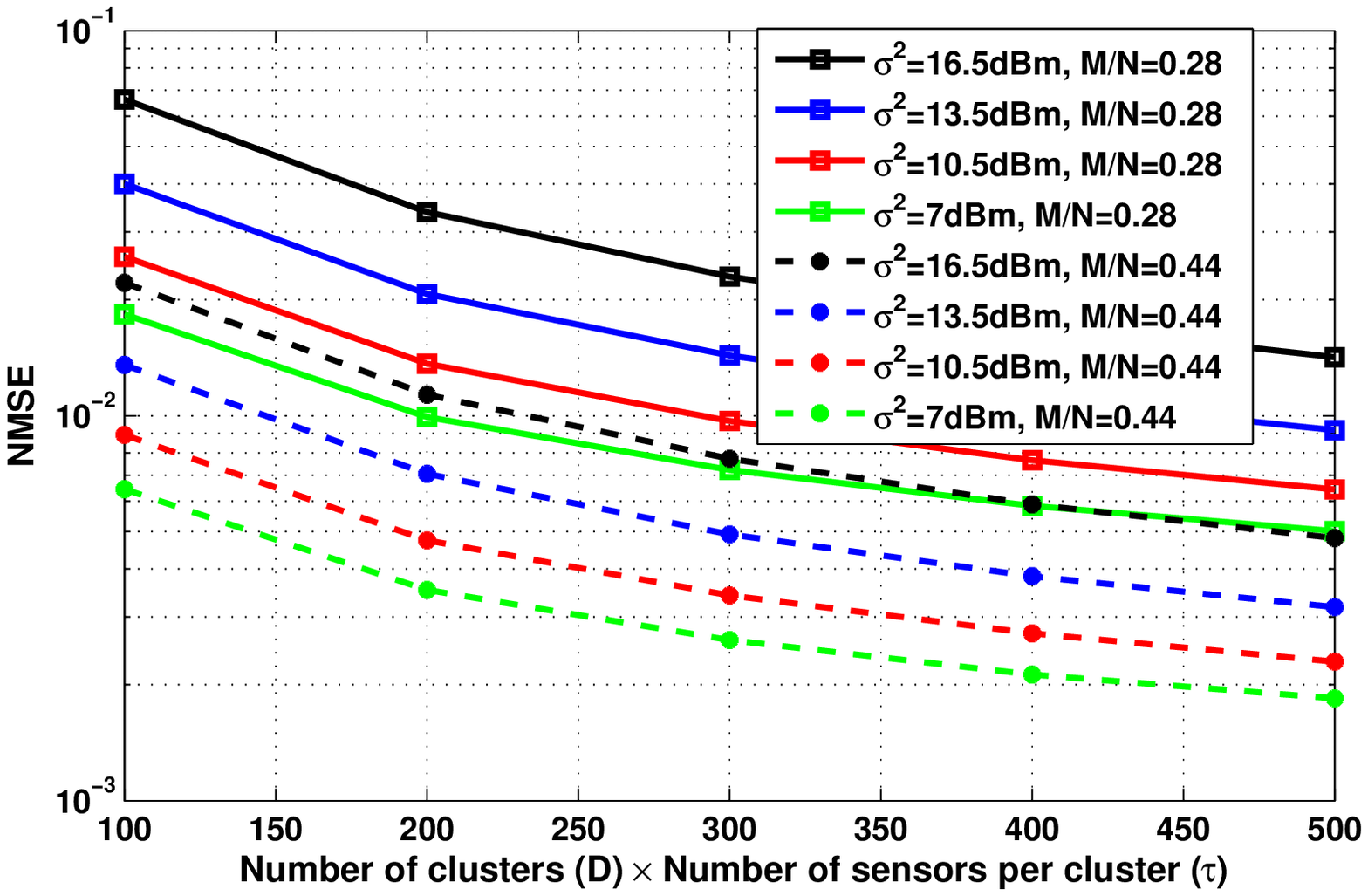}\vspace{-1.5mm}
        \caption{The NMSE between the 
        CAP and the NAP for the second experiment (synchronized sensors).}        
        \label{fig:NMSEExperiment2}
\end{figure}
The second experiment uses the same setting as used in the first experiment (including Table~\ref{tab:experiment1}). The only difference is that the 
sensors are now assumed to be synchronized. Fig.~\ref{fig:DisplayExperiment2} depicts the CAP and the NAP of the faded user signals received at the 
sensors. 
{\color{blue}Unlike in the unsynchronized sensors case (see Fig.~\ref{fig:DisplayExperiment1}), we now observe a significant variation in both the CAP and the NAP. This is because, when the sensors are synchronized, they observe the same part of the user signals. 
This means that, while the fading realization components in the received signals at different sensors are independent, the user signal components in the received signals at different sensors are fully correlated.} 
Fig.~\ref{fig:NMSEExperiment2} shows the NMSE of the CAP with respect to the NAP for the synchronized sensors case. In general, some trends found in the unsynchronized sensors case also appear here. Notice that the NMSE 
for the synchronized sensors case is smaller than the one for the unsynchronized sensors case since the quality of the NAP in the synchronized sensors case is also significantly worse than the one in the unsynchronized sensors case. Note that we can also re-interpret this second experiment for problem P1. This re-interpretation however, will make more sense, if we reverse the roles of $H_t^{(k)}(\vartheta)$ and $U_t^{(k)}(\vartheta)$. When this is the case, for P1, the second experiment implies that each user transmits temporally independent signals and that the signals from different users $k$ pass through statistically different and uncorrelated {\it time-invariant} fading channels 
on their way towards the receiving array. Here, the statistics of the user signal are constant only within a cluster of time indices.

\begin{figure}[h]
\begin{minipage}[b]{1\linewidth}
  \centering
  \includegraphics[width=1\textwidth]{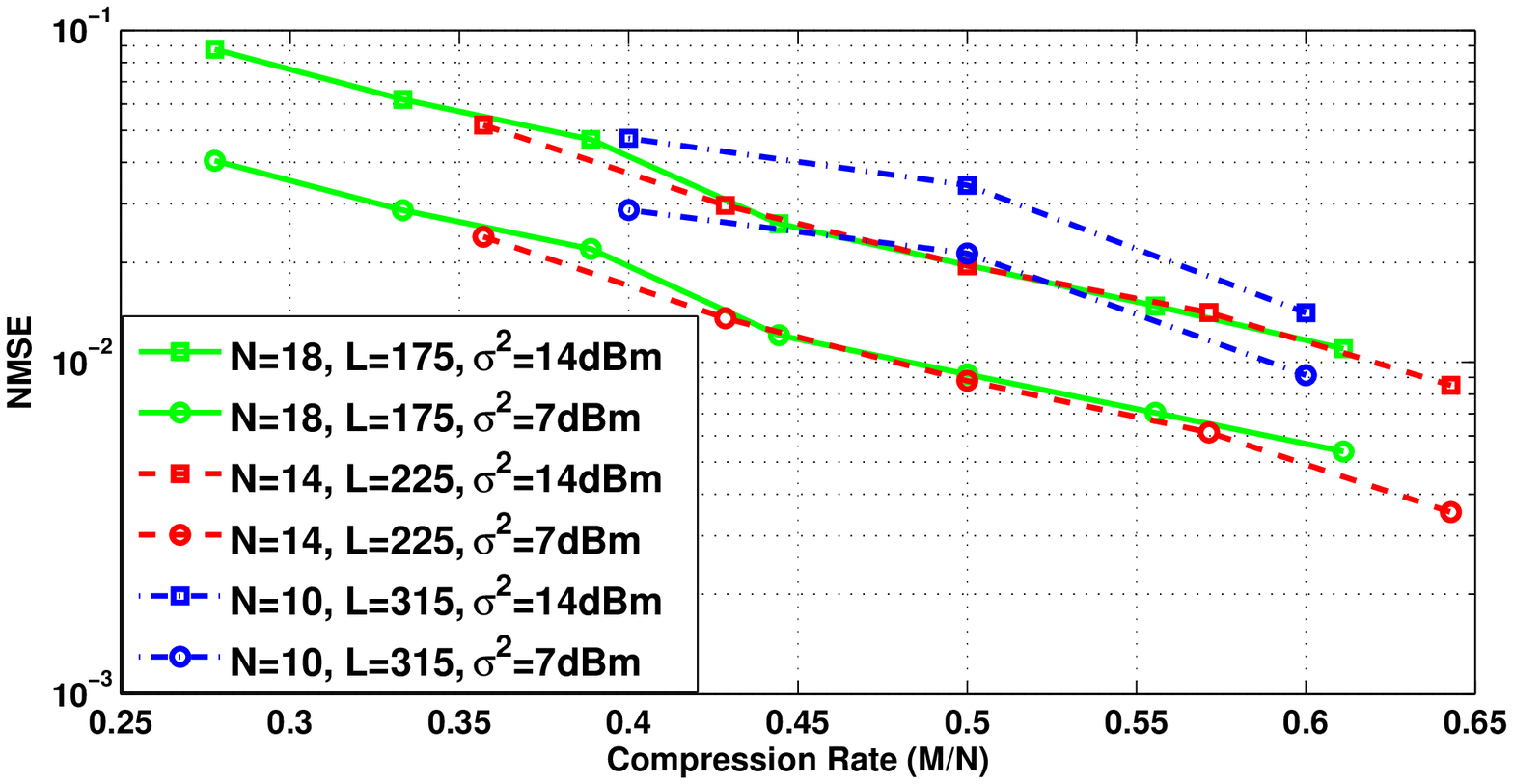}
  \centerline{\small(a)}
\end{minipage}
\begin{minipage}[b]{1\linewidth}
  \centering
  \includegraphics[width=1\textwidth]{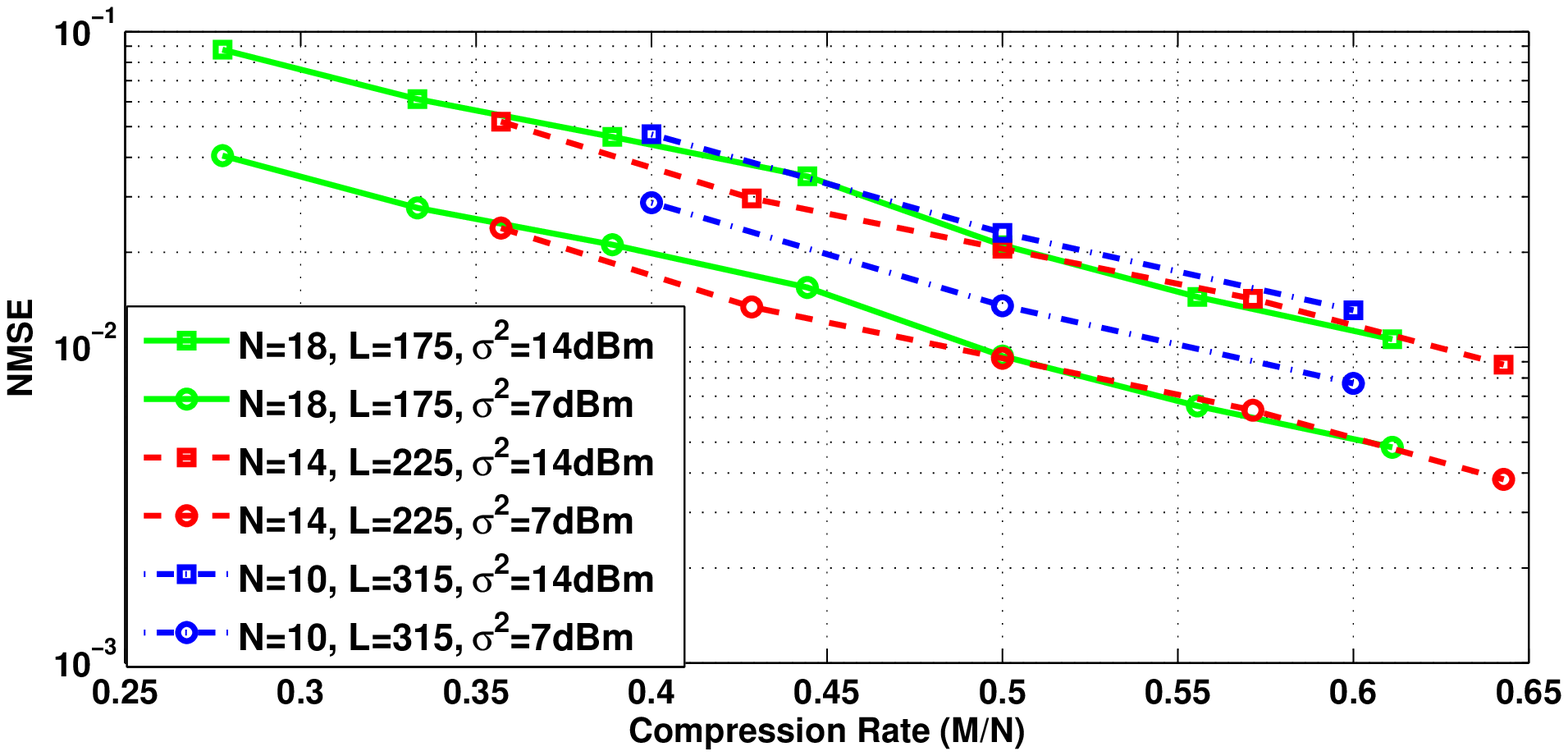}
  \centerline{\small(b)}
\end{minipage}
        \caption{The NMSE between the CAP and the NAP for the third experiment (comparison of different bin size); (a) using the first set of coset patterns (see Table~\ref{tab:coset2sets}); (b) using the second set of coset patterns.}        
        \label{fig:NMSEDiffBinSize}
\end{figure}

\begin{table}[th]
	\caption{The two sets of coset patterns used in the third experiment (comparison of different bin size).}
	\centering
	\vspace{-1mm}
		\begin{tabular}{| c | c | c |}
			\hline
			\multicolumn{3}{|c|}{First set of coset patterns}\\
			\hline
    $N$ & Minimal circular & The order of the additional coset indices\\  
		  & sparse ruler indices & for implementing a larger compression rate\\ \hline
   18 & 0, 1, 4, 7, 9 & 17, 2, 13, 12, 15, 6 \\ \hline
   14 & 0, 1, 2, 4, 7 & 10, 6, 12, 5 \\ \hline
   10 & 0, 1, 3, 5 & 8, 4 \\ \hline
			\multicolumn{3}{|c|}{Second set of coset patterns}\\
			\hline
    $N$ & Minimal circular & The order of the additional coset indices\\  
		  & sparse ruler indices & for implementing a larger compression rate\\ \hline
   18 & 0, 1, 4, 7, 9 & 5, 2, 6, 17, 15, 14  \\ \hline
   14 & 0, 1, 2, 4, 7 & 12, 10, 13, 11 \\ \hline
   10 & 0, 1, 3, 5 & 4, 6 \\ \hline
    \end{tabular}
\label{tab:coset2sets}
\end{table}
{\color{blue}In the third experiment, we investigate the impact of varying the bin size (which is equivalent to varying $N$) and $L$ for a given $\tilde{N}$ on the performance of the CAP approach. Let us consider 
the settings in the first experiment (i.e., we consider Table~\ref{tab:experiment1}) 
except for the following. 
We now examine three different values of $N$, i.e., $N=10$, $N=14$, and $N=18$ for a given $\tilde{N}=3150$. For each value of $N$, we vary the compression rate $M/N$ and examine the two sets of coset patterns available in Table~\ref{tab:coset2sets}. We start from the minimal 
$M/N$ offered by the minimal circular sparse ruler. Larger compression rates are implemented by selecting additional coset indices where the order of the selection is provided by the third column of Table~\ref{tab:coset2sets}. 
We fix the number of $\tau$ 
to $\tau=76$ and perform 1000 Monte Carlo simulation runs for different noise variances (see Fig.~\ref{fig:NMSEDiffBinSize}).
Fig.~\ref{fig:NMSEDiffBinSize} illustrates the NMSE of the CAP with respect to the NAP for the two sets of coset patterns. Observe that varying $N$ and $L$ for a given $\tilde{N}$ does not really result in a clear trend in the estimation performance. While the performance of the CAP for $N=10$ is worse than the one for the larger value of $N$, the performance of the CAP for $N=14$ is better than the one for $N=18$ for some values of $M/N$. Note that the NMSE also depends on the coset pattern that we select to implement a particular compression. 
At this point, we would like to mention that, as long as the bin size constraint in Remark~2 is satisfied, having a larger $N$ is generally more advantageous as we will generally have a lower value of minimum $M/N$. This is because it can be found that, as $N$ increases, the number of marks in the corresponding length-$(N-1)$ minimal circular sparse ruler (which is the minimum $M$) tends to be constant or to increase very slowly. As a result, 
the minimum compression rate $M/N$ also generally (even though not monotonically) decreases with $N$.} 

\begin{figure}[t]
				\centering
        \includegraphics[width=0.49\textwidth]{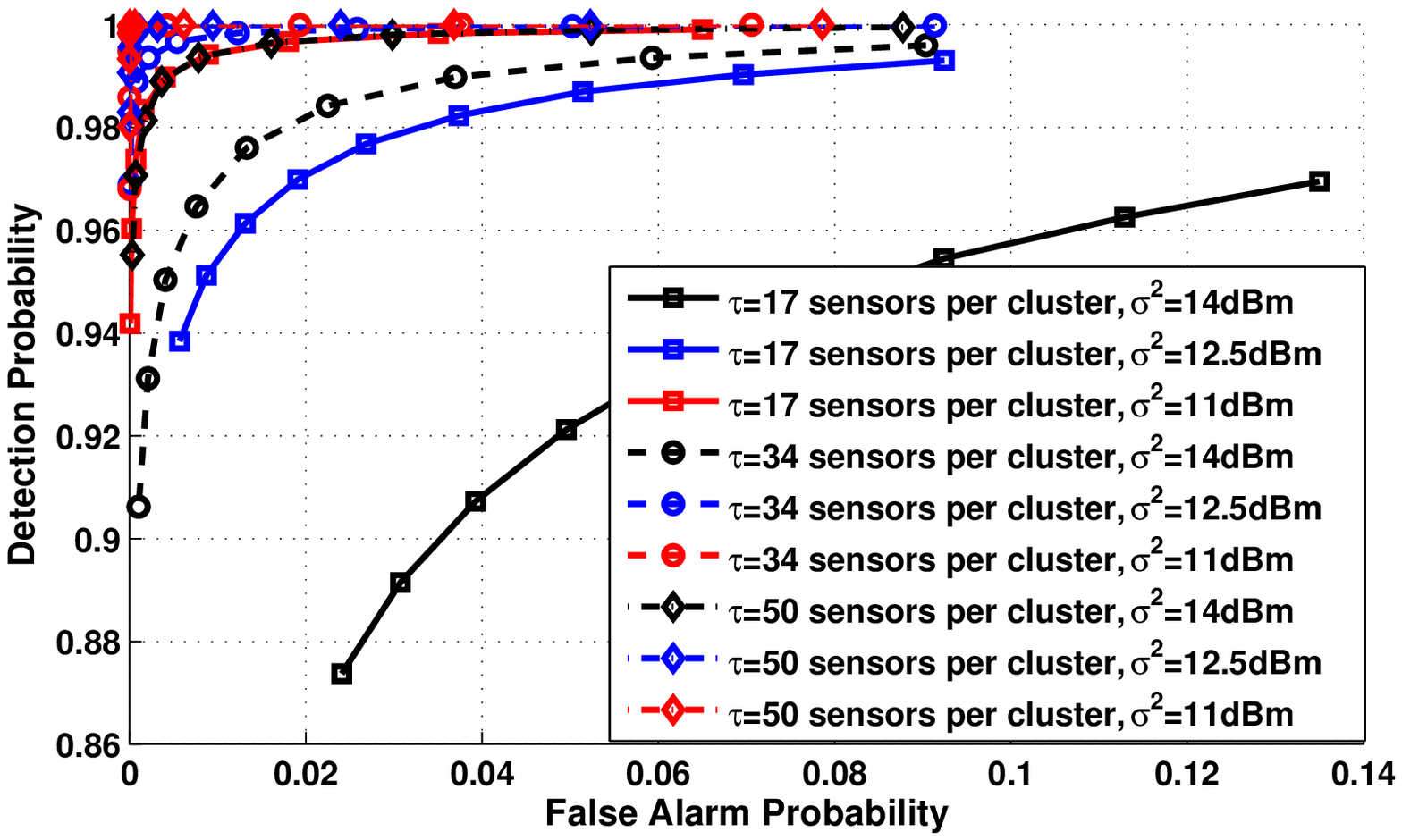}\vspace{-2mm}
        \caption{The resulting ROC when the 
        CAP is used to detect the existence of the active user signals suffering from fading channels in the fourth experiment (unsynchronized sensors).}        
        \label{fig:DetectSc1Op3}
				\centering
        \includegraphics[width=0.49\textwidth]{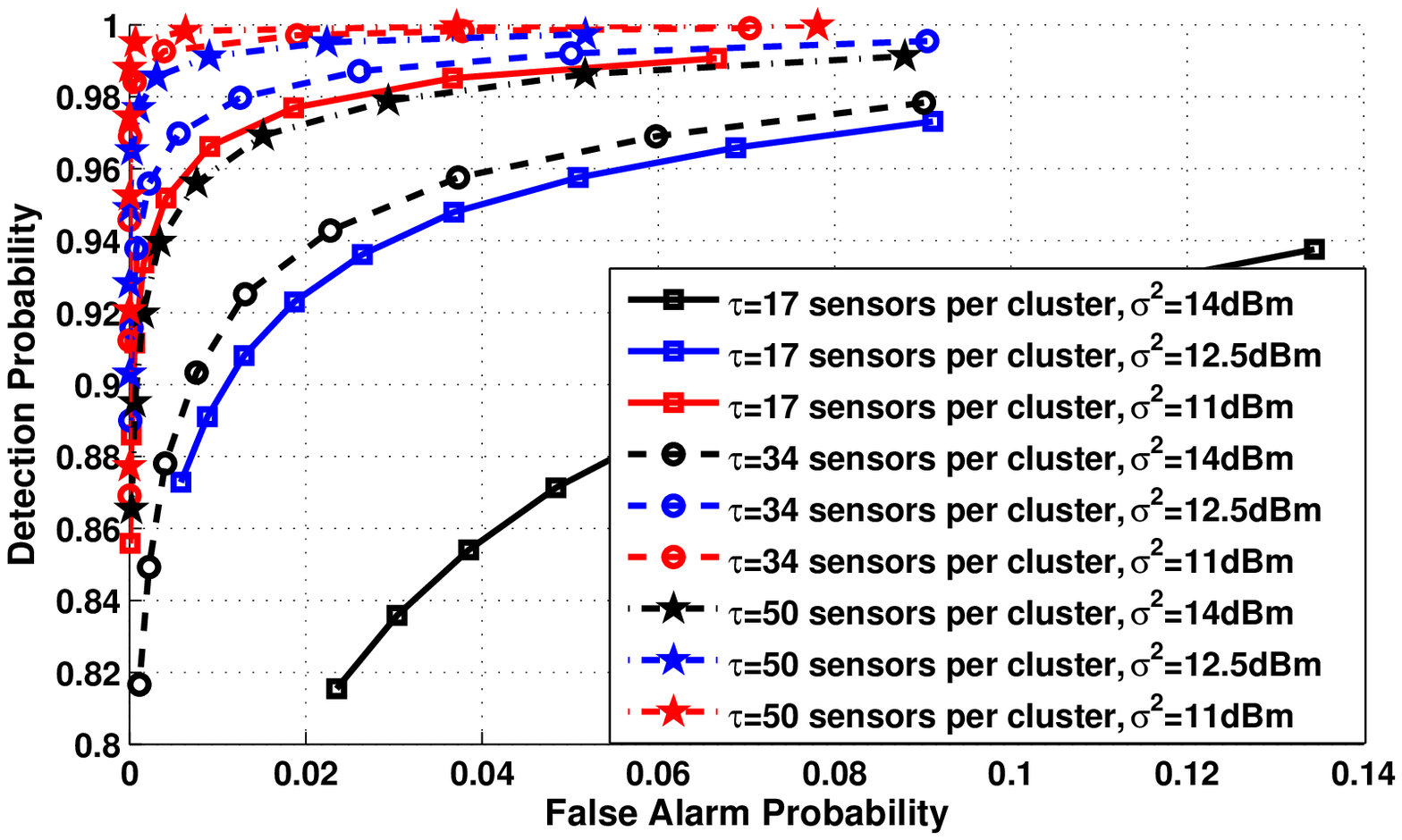}\vspace{-2mm}
        \caption{The resulting ROC when the 
        CAP is used to detect the existence of the active user signals suffering from fading channels in the fifth experiment (synchronized sensors).}        
        \label{fig:DetectSc1OpTwo}
\end{figure}

In the next {\color{blue}three} experiments, we use the 
CAP to detect the existence of active user signals that suffer from fading channels and evaluate the detection performance. We start with the fourth experiment, where we again consider problem P2, $\tilde{N}=3060$, $L=170$, $N=18$, and $M/N=0.28$ (again by adopting 
$\mathcal{M}=\{0,1,4,7,9\}$). We now consider $D=3$ clusters of $\tau$ unsynchronized sensors and $K=3$ user signals (see their settings in Table~\ref{tab:experimentdetect}), which 
are generated using the same procedure used in the first experiment. 
The amount of path loss (which includes shadowing) experienced between each user and each cluster is listed in Table~\ref{tab:experimentdetect}. We then simulate a small-scale Rayleigh fading channel on top of it. We perform 5000 Monte Carlo runs and vary 
$\tau$ and 
$\sigma^2$ 
(see Fig.~\ref{fig:DetectSc1Op3}). 
We vary the detection threshold manually and out of the $\tilde{N}=3060$ frequency points at which the CAP is reconstructed, we evaluate the resulting detection events at $363$ frequency points in the active bands and the false alarm events at $363$ frequency points in the bands that are far from the active bands, i.e., $[-0.77\pi,-0.53\pi]$. 
Here, we average the estimated power over every eleven subsequent frequency points $\vartheta$ and apply the threshold to these average values. The resulting receiver operating characteristic (ROC) is depicted in Fig.~\ref{fig:DetectSc1Op3}. Observe the acceptable detection performance of the CAP 
for the examined $\tau$ and $\sigma^2$ though the performance is slightly poor for $\tau=17$ and $\sigma^2=14$ dBm. 
This detection performance demonstrates that the proposed CAP 
can be used in a spectrum sensing application such as in a CR network.
\begin{table}[ht]
	\caption{The frequency band and the power of the user signals and the experienced path loss in the fourth and the fifth experiments.}
	\centering
	\vspace{-1mm}
		\begin{tabular}{| c | c | c | c | c |}
			\hline
    User band & Power/freq. & \multicolumn{3}{c|}{Path loss (in dB) at cluster}\\ \cline{3-5} 
		 (rad/sample) &(per rad/sample)& 1 & 2 & 3 \\ \hline
    $[0.41\pi,0.49\pi]$ & $25$ dBm& $-12$ & $-13$ & $-14$ \\ \hline
    $[0.31\pi,0.39\pi]$ & $25$ dBm& $-14.5$ & $-13$ & $-11.5$ \\ \hline
    $[0.21\pi,0.29\pi]$ & $25$ dBm& $-13.5$ & $-13$ & $-12.5$ \\ \hline
    \end{tabular}
\label{tab:experimentdetect}\vspace{-2mm}
\end{table}
\begin{figure}[h]
				\centering
        \includegraphics[width=0.49\textwidth]{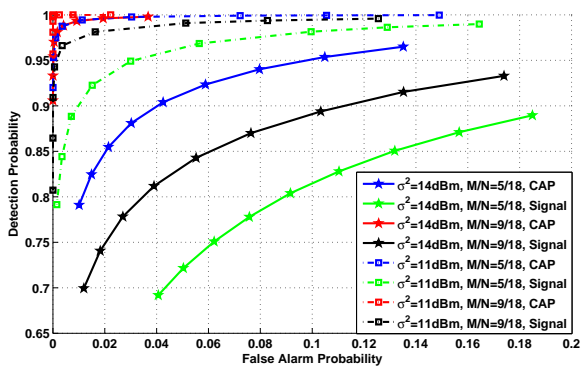}\vspace{-2mm}
        \caption{The resulting ROC of the detector when the CAP is used compared with the one when the compressive signal reconstruction using RM-FOCUSS of~\cite{BhaskarRao} is used (the sixth experiment).}        
        \label{fig:CompareMMV}
\end{figure}

The fifth experiment repeats the fourth experiment but for synchronized sensors. 
The ROC in Fig.~\ref{fig:DetectSc1OpTwo} shows that the detection performance for the synchronized sensors case is 
worse than the one for the unsynchronized sensors case in Fig.~\ref{fig:DetectSc1Op3} 
due to the significant variation in the 
CAP as shown in Fig.~\ref{fig:DisplayExperiment2}.

{\color{blue}In the sixth experiment, we consider problem P2 and compare the detection performance of the spectrum sensing approach based on CAP with that of the one based on the RM-FOCUSS discussed in Section~\ref{complexity}.
To simulate the existence of a joint sparsity structure in $\{{\bf x}_{t}(\vartheta)\}_{t=1}^{\tau}$, we again use the settings 
in Table~\ref{tab:experimentdetect}. However, we now only assume one cluster of $\tau=50$ sensors where the amount of path loss experienced between each user and each sensor is set to $-13$ dB. 
The ROC for 5000 Monte Carlo runs and different $M/N$ as well as $\sigma^2$ is illustrated in
Fig.~\ref{fig:CompareMMV}. Here, the compression rate of $M/N = 9/18$ is implemented by activating four extra cosets, i.e., {16, 8, 12, 13} (which we decide randomly), on top of the length-$17$ minimal circular sparse ruler. The M-FOCUSS convergence criterion parameter 
and the M-FOCUSS diversity measure parameter (labeled as $p$ in~\cite{BhaskarRao}) are set to $0.001$ and $0.8$, respectively. Note that the latter setting follows the suggestion of~\cite{BhaskarRao}. To determine the M-FOCUSS regularization parameter, 
we first perform some experiments and examine ten different values of regularization parameters between $10^{-4}$ and $10$. 
We then select the regularization parameter that leads to the smallest NMSE between the resulting compressive estimate of $\{|{X}_{t}(\vartheta)|^2\}_{t=1}^{\tau}$, for all the considered $\vartheta\in[0,1)$, and the Nyquist-rate version. 
We finally decide to set the regularization parameter to $10$ for the case of $M/N=5/18$, to $0.01668$ for the case of $M/N=9/18$ and $\sigma^2=14$ dBm, and to $0.21544$ for the case of $M/N=9/18$ and $\sigma^2=11$ dBm (see Fig.~\ref{fig:CompareMMV}). Observe from Fig.~\ref{fig:CompareMMV} that the spectrum sensing approach based on CAP has a better detection performance than the one based on signal/spectrum reconstruction using RM-FOCUSS. 
Recall that the approach of~\cite{BhaskarRao} requires the sparsity constraint on the vectors to be reconstructed (which are $\{{\bf x}_{t}(\vartheta)\}_{t=1}^{\tau}$). This implies that, if we have additional active users on top of the scenario used in the sixth experiment, the actual $\{{\bf x}_{t}(\vartheta)\}_{t=1}^{\tau}$ will have a smaller sparsity level. In this case, if we use the same compression rate $M/N$ as the one used in the sixth experiment, the performance of RM-FOCUSS will be even worse.} 

\subsection{Correlated Bins}\label{simulation_corrbins}

\begin{figure}[t]
				\centering
        \includegraphics[width=0.49\textwidth]{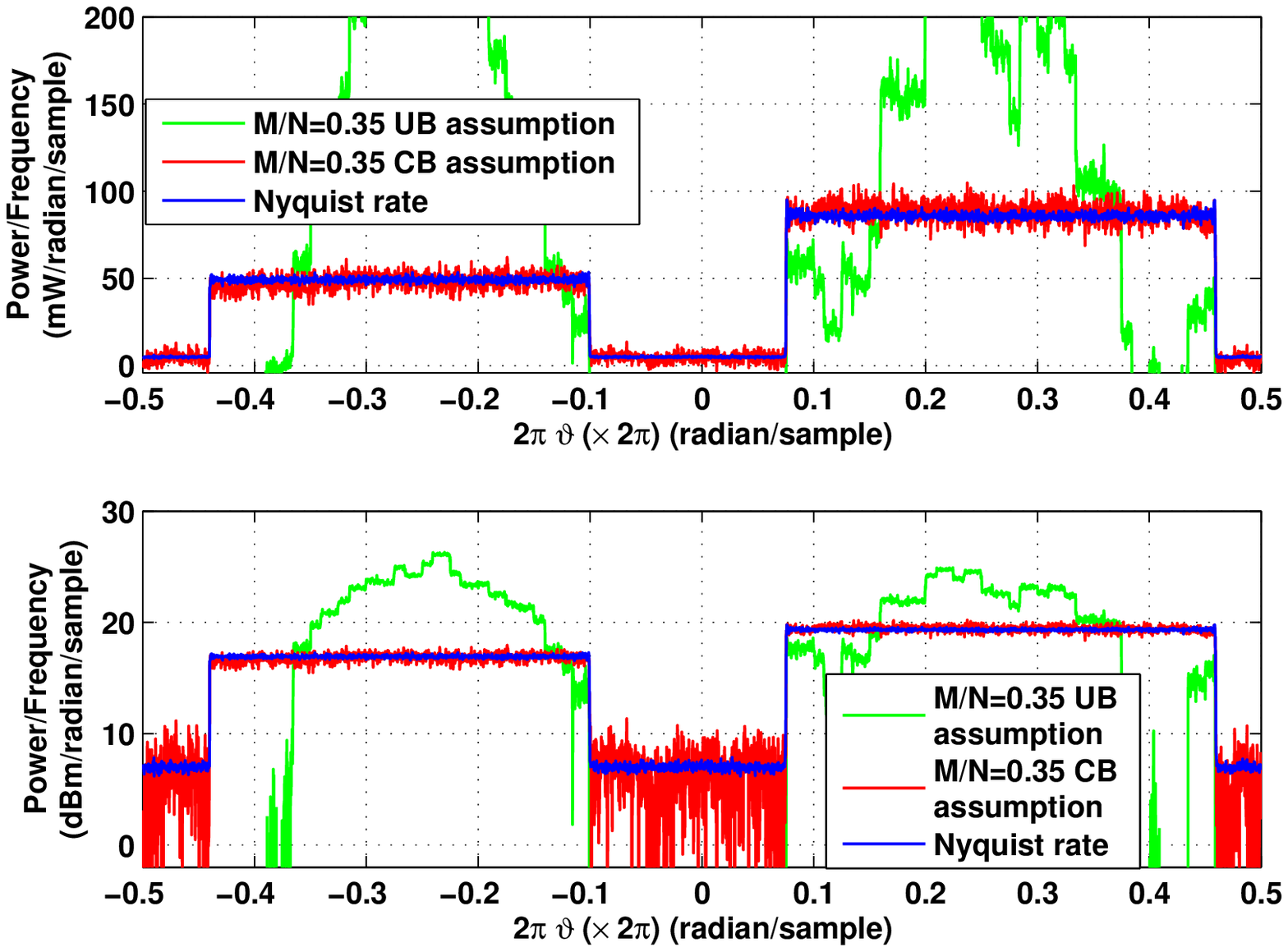}\vspace{-2.5mm}
        \caption{The 
        CAP and the NAP of the faded user signals for the seventh experiment in Section~\ref{simulation_corrbins} as a function of frequency in a linear scale (top) and logarithmic scale (bottom).}        
        \label{fig:DisplayCorrelated}\vspace{0.5mm}
				\centering
        \includegraphics[width=0.49\textwidth]{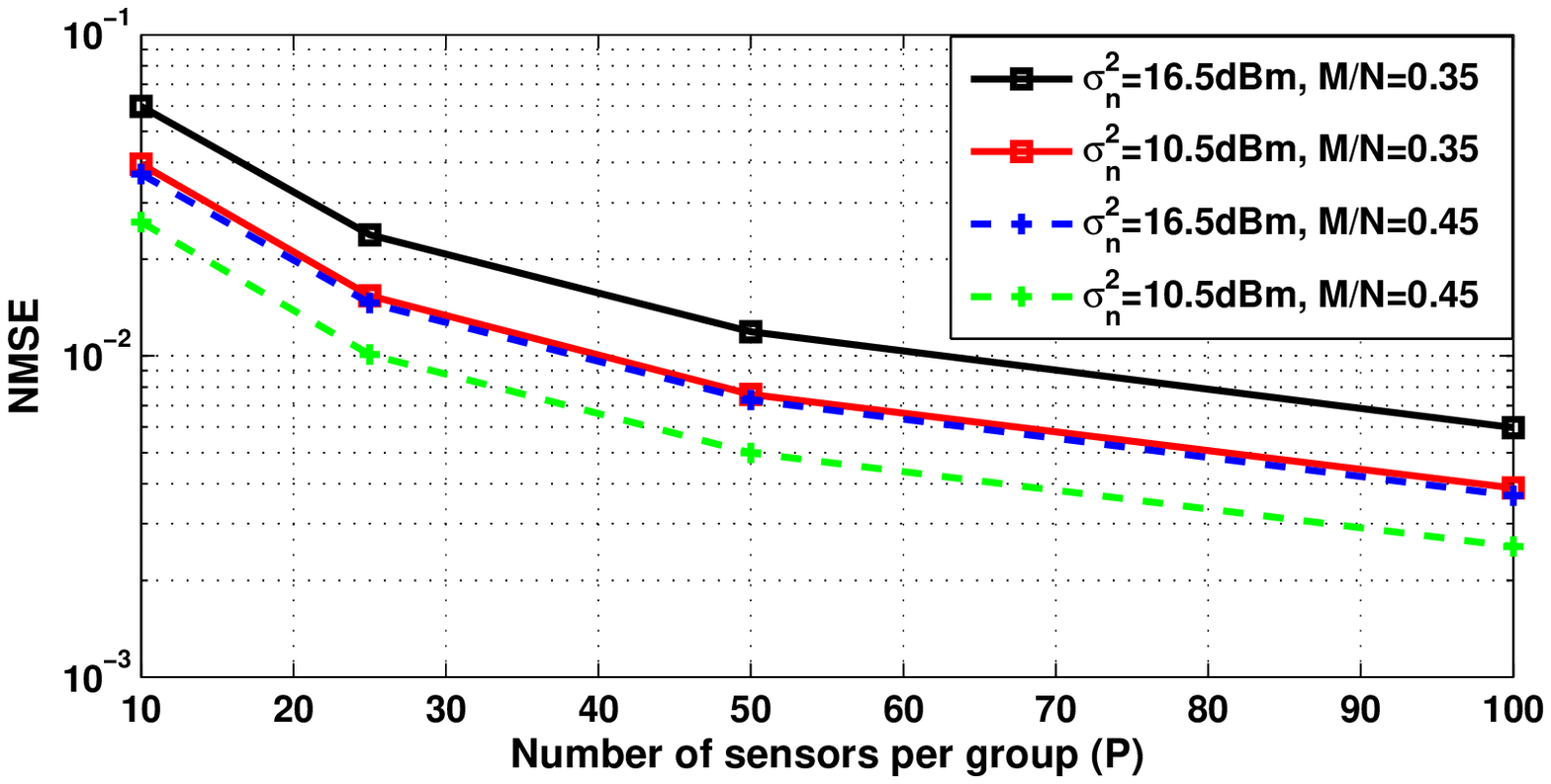}\vspace{-2mm}
        \caption{The NMSE between the 
        CAP based on the correlated bins assumption and the 
        NAP for the seventh experiment in Section~\ref{simulation_corrbins}.}        
        \label{fig:NMSECorrelated}
\end{figure}

In this section, we conduct the seventh experiment to evaluate the estimation performance of the 
CAP approach for the correlated bins case discussed in Section~\ref{correlatedbins}. 
Here, we consider problem P2, $\tilde{N}=3080$, $L=77$, $N=40$, and $M=14$ ($M/N=0.35$). Recall from Section~\ref{correlatedbins} that the mathematical model for the correlated bins case is similar to the one in~\cite{ElsevierDOA}. Hence, to design the sampling matrices for all sensors, which are assumed to be synchronized, that ensure the full column rank 
of ${\boldsymbol \Psi}$ in~\eqref{eq:Psi}, we use the algorithm of~\cite{ElsevierDOA}, which is originally designed to solve the antenna selection problem for estimating the DOA of highly correlated sources. This algorithm, which only offers a suboptimal solution for 
$Z$, suggests 
$Z=12$ groups of $P=25$ sensors where each group has a unique set of $M=14$ active cosets. 
We consider $K=2$ user signals whose setting is given in Table~\ref{tab:experiment5corr}. 
To simulate the full correlation between all the frequency components within the band of the $k$-th user, we assume that the $k$-th user transmits exactly the same symbol at all these frequency components at each time instant. 
On its way toward the different sensors, the signal of the $k$-th user is assumed to pass through different and uncorrelated Rayleigh fading channels $H_t^{(k)}(\vartheta)$ but it suffers from the same path loss and shadowing, whose value is 
listed in Table~\ref{tab:experiment5corr}. 
Again, we 
assume flat fading in each user band and have $\sigma^2=7$ dBm.
Fig.~\ref{fig:DisplayCorrelated} shows the 
CAP of the faded user signals using the correlated bins (CB) assumption. As a benchmark, we also provide 
the NAP and the 
CAP based on the uncorrelated bins (UB) assumption discussed in Sections~\ref{uncorr_bins_system_model}-\ref{performance}, which 
is obtained by activating the same set of $M=14$ cosets, i.e., $\mathcal{M}=\{0,1,2,3,4,9,10,15,16,18,20,30,33,37\}$, in all sensors
(leading to a full column rank matrix ${\bf R}_c$ in~\eqref{eq:Rybar_as_rbar_x}). Observe that the quality of the 
CAP based on the UB assumption is extremely poor. 
On the other hand, with respect to the NAP, the degradation in the quality of the 
CAP based on the CB assumption is acceptable despite a significant variation 
in the unoccupied band. 
Next, we perform 1000 Monte Carlo runs and vary the number of sensors per group $P$, 
$\sigma^2$, 
and 
$M/N$ (see Fig.~\ref{fig:NMSECorrelated}). 
In Fig.~\ref{fig:NMSECorrelated}, the compression rate of $M/N=0.45$ is implemented by randomly activating four additional cosets on top of the already selected $14$ cosets and the resulting sampling pattern is kept fixed throughout the entire Monte Carlo runs. 
Fig.~\ref{fig:NMSECorrelated} shows the NMSE of the 
CAP based on the CB assumption with respect to the NAP, which indicates that either increasing $M/N$ 
or having more sensors per group $P$ can significantly improve the estimation quality. Again, 
a larger NMSE is introduced for a larger noise power.
\begin{table}[t]
	\caption{The frequency bands occupied by the users, their power, and the experienced path loss in the seventh experiment.}
	\centering
	\vspace{-1mm}
		\begin{tabular}{| c | c | c |}
			\hline
    User band & Power/freq. &Path loss\\ 
		 (rad/sample) &(per rad/sample)&\\ \hline
    $[-0.88\pi,-0.2\pi]$ & $22$ dBm& $-6$ dB\\ \hline
    $[0.15\pi,0.92\pi]$ & $25$ dBm& $-7$ dB\\ \hline
    \end{tabular}
\label{tab:experiment5corr}
\end{table}

The interpretation of this seventh experiment for P1 is similar to the problem discussed in~\cite{ElsevierDOA}. For P1, this experiment is equivalent to having a ULA consisting of $N=40$ ULSs, where the 
array scanning time is split into $P=25$ scanning periods, each of which consists of $Z=12$ time slots. In different time slots per scanning period, we activate different sets of $M=14$ (out of $N=40$) ULSs 
leading to a DLA. 
The interpretation will again make more sense if we reverse the roles of $H_t^{(k)}(\vartheta)$ and $U_t^{(k)}(\vartheta)$. When this is the case, for P1, the experiment implies that all users transmit temporally independent signals and that the signals from different users $k$ pass through statistically different and uncorrelated time-invariant fading channels 
on their way towards the receiving array.
As the signal received from the $k$-th user at different angles within its angular band is fully correlated, this can be related to a situation 
where the same symbol of the $k$-th user hits different scatterers (which play the role of the channel) before reaching the observing array. From the point of view of the array, the scattered versions of the symbol will be received from different angles within a particular angular band.

\begin{figure}[h]
				\centering
        \includegraphics[width=0.49\textwidth]{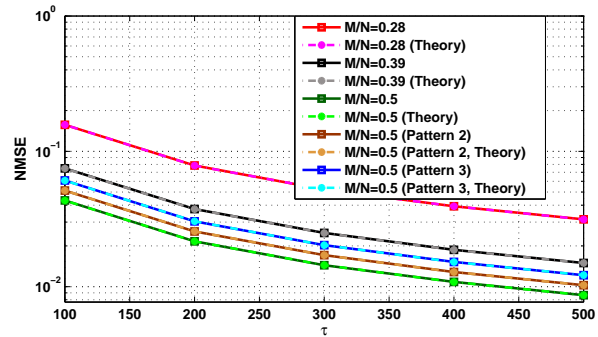}\vspace{-3mm}
        \caption{The simulated and analytical NMSE between the 
        CAP and the true power spectrum 
        when $x_t[\tilde{n}]$ only contains circular complex Gaussian i.i.d. noise. 
Unless mentioned otherwise, the cases of $M/N>0.28$ are implemented by activating extra cosets based on Pattern~1.}        
        \label{fig:NMSEGaussianNoise}
\end{figure}
\begin{table}[th]
	\caption{Three coset patterns to be added on top of the already selected minimal circular sparse ruler based coset indices for implementing $M/N>0.28$
	in Section~\ref{simulation_noise}.}
	\centering
	\vspace{-1mm}
		\begin{tabular}{| c | c |}
			\hline
    Coset pattern & The order of the additional coset indices\\ \hline  
    Pattern 1 & 17, 11, 2, 6 \\ \hline
    Pattern 2 & 3, 5, 6, 8 \\ \hline
    Pattern 3 & 2, 3, 5, 6 \\ \hline
    \end{tabular}
\label{tab:cosetpattern}
\end{table}
\subsection{Circular Complex Gaussian Noise}\label{simulation_noise}
The last experiment examines the performance of the 
CAP based on the UB assumption when the received signal $x_t[\tilde{n}]$ only contains circular complex zero-mean Gaussian spatially and temporally i.i.d. noise. Here, we have $\tilde{N}=3060$, $L=170$, $N=18$, and $\sigma^2=7$ dBm. We perform 1000 Monte Carlo 
runs and vary $\tau$ (see Fig.~\ref{fig:NMSEGaussianNoise}). 
We compute the NMSE of the 
CAP with respect to the true power spectrum (since $x_t[\tilde{n}]$ in this case is clearly a WSS signal) and compare this NMSE obtained from the simulation with the analytical NMSE. Since it can be 
shown that, for circular complex Gaussian i.i.d. noise $x_t[\tilde{n}]$, $\hat{P}_{{x},{LS}}(\vartheta)$ is an unbiased estimate of ${P}_x(\vartheta)$ even for finite $\tilde{N}$, the analytical NMSE only depends on the variance of $\hat{P}_{{x},{LS}}(\vartheta)$ and it can be shown to be equal to $\frac{1}{\tau}(\frac{1}{M}+\sum_{n=1}^{N-1}\frac{1}{\gamma_{n+1}})$ by using~\eqref{eq:VarPxLSwhitenoisetheo}. We start with $M/N=0.28$ by using the cosets indexed 
by the length-$17$ minimal circular sparse ruler, i.e., $\mathcal{M}=\{0,1,4,7,9\}$, and then vary $M/N$. 
First, the cases of $M/N>0.28$ are implemented by activating additional cosets based on Pattern~1 in Table~\ref{tab:cosetpattern}. Then, we also test Pattern~2 and Pattern~3 as additional coset patterns to implement the case of $M/N=0.5$. 
Observe in Fig.~\ref{fig:NMSEGaussianNoise} how the analytical NMSE is on top of the simulated NMSE for all the evaluated $M/N$ values. 
Also observe that, for $M/N=0.5$, the three different coset patterns have led to different values of the NMSE depending on the resulting value of $\{\gamma_{n+1}\}_{n=1}^{N-1}$ in~\eqref{eq:VarPxLSwhitenoisetheo}. 

\section{Conclusion and Future Work}\label{sec:conclusion}

This paper proposed a compressive periodogram reconstruction approach and considered both 
time-frequency and spatio-angular domains. In our model, the entire band is split into uniform bins 
such that the received spectra at two frequencies or angles, whose distance is equal to or larger than the size of a bin, are uncorrelated. 
In both considered domains, this model leads to a 
circulant coset correlation matrix, which allows us to perform a strong compression yet to present our reconstruction problem as an overdetermined system. 
When the 
coset patterns are designed based on a circular sparse ruler, the system matrix has full column rank and we can reconstruct the periodogram using LS. 
In a practical situation, our estimate of the coset correlation matrix is only asymptotically circulant. 
Hence, we also presented an asymptotic bias and variance analysis for the 
CAP. We further included 
a thorough variance analysis on 
the case when the received signal only contains circular complex zero-mean white Gaussian noise, which 
provides some useful insights in the performance of our 
approach. 
The variance analysis for a more general signal (i.e., a general Gaussian signal) has also been presented but it is not easy to interpret due to its dependence on the unknown statistics of the user signals.
We also proposed a solution for the case when the bin size is decreased such that the received spectra at two frequencies or angles, with a spacing
between them larger than the size of the bin, can still be correlated. 
Finally, the simulation study showed that the estimation performance of the evaluated approach is acceptable 
and that our CAP 
performs well when detecting the existence of the user signals suffering from 
fading channels. 

{\color{blue}As a future work, we are interested in the case when both problems P1 and P2 emerge simultaneously. In that case, we would consider a compressive linear array of antennas 
and a compressive digital recever unit per antenna 
leading to a two-dimensional (2D) digital signal. Our interest would then be to investigate if it is possible to perform 
compression in both the time and spatial domain and to jointly reconstruct the angular and frequency periodogram from the 2D compressive samples. To study that, we could follow an approach similar 
to~\cite{EUSIPCO}, which assumes stationarity in both the time and spatial domain and exploits the existing Toeplitz structure in the correlation matrix.}

\appendices
\section{Proof of Theorem~2}
\label{ProofTheorem2}
Recall that $\hat{\bf R}_{{\bar{y}}}(\vartheta)$ in~\eqref{eq:Rybar_hat} is an unbiased estimate of ${\bf R}_{\bar{y}}(\vartheta)$ in~\eqref{eq:Ry_bar}, i.e., $E[\hat{\bf R}_{{\bar{y}}}(\vartheta)]={\bf R}_{\bar{y}}(\vartheta)$. 
Applying the expectation operator on~\eqref{rxhatbarLS} and~\eqref{eq:RxLS}, it is then clear that $\hat{\bf r}_{\bar{x},LS}(\vartheta)$ in~\eqref{rxhatbarLS} and $\hat{\bf R}_{x,LS}(\vartheta)$ in~\eqref{eq:RxLS} are unbiased estimates of ${\bf r}_{\bar{x}}(\vartheta)$ in~\eqref{eq:Rybar_as_rbar_x} and ${\bf R}_{x}(\vartheta)$ in~\eqref{eq:Rxt_bar}, respectively, since ${\bf r}_{\bar{x}}(\vartheta)$ in~\eqref{eq:Rybar_as_rbar_x} can perfectly be reconstructed from ${\bf R}_{\bar{y}}(\vartheta)$ using LS. Recall from Remark~1 that the $(i+1)$-th diagonal element of ${\bf R}_{x}(\vartheta)$ is equal to $E[|X_{t,i}(\vartheta)|^2]$. From~\eqref{CRAP}, it is then obvious that the CAP $\hat{P}_{x,LS}(\vartheta+\frac{i}{N})$ is an unbiased estimate of $\frac{1}{\tilde{N}}E[|X_{t,i}(\vartheta)|^2]$. However, by taking~\eqref{eq:PowerSpectrum} into account, we can observe that 
\vspace{-1mm}
\begin{equation}
\lim_{\tilde{N}\rightarrow\infty}\frac{1}{\tilde{N}}E[|X_{t,i}(\vartheta)|^2]=P_x(\vartheta+\frac{i}{N}),\:\:\vartheta\in[0,1/N), 
\label{AsymptoticEXtivartheta}
\vspace{-1mm}
\end{equation}
for $i=0,1,\dots,N-1$, since $x_t[\tilde{n}]$ is a finite-length observation of the actual random process $x[\tilde{n}]$. Hence, by applying $\lim_{\tilde{N}\rightarrow\infty}E[\hat{P}_{x,LS}(\vartheta+\frac{i}{N})]$ and using~\eqref{AsymptoticEXtivartheta}, it is clear that $\hat{P}_{x,LS}(\vartheta+\frac{i}{N})$ is an asymptotically (with respect to $\tilde{N}$) unbiased estimate of $P_x(\vartheta+\frac{i}{N})$ in~\eqref{eq:PowerSpectrum}, for $\vartheta \in [0,1/N)$ and $i=0,1,\dots,N-1$. $\square$

\vspace{-2mm}
\section{Proof of Proposition~1}
\label{ProofPropos1}

Note that for the specific case assumed in this proposition, we can rewrite~\eqref{eq:CovRycheckbarelementGaussian} as 
\vspace{-1mm}
\begin{align}
&\text{Cov}[[\hat{\bf R}_{{\bar{y}}}(\vartheta)]_{m+1,m'+1},[\hat{\bf R}_{{\bar{y}}}(\vartheta)]_{a+1,a'+1}]\nonumber\\
&=\frac{1}{N^4\tau^2}\sum_{t=1}^{\tau}\sum_{i=0}^{N-1}\sum_{i'=0}^{N-1}\sum_{b=0}^{N-1}\sum_{b'=0}^{N-1}e^{\frac{j2\pi (n_m i-n_{m'} i'-n_ab+n_{a'}b')}{N}}\times\nonumber\\
&E[{X}_{t,i}(\vartheta){X}_{t,b}^*(\vartheta)]E[{X}_{t,i'}^*(\vartheta){X}_{t,b'}(\vartheta)],
\label{eq:CovRycheckbarelementGaussian_noise}
\vspace{-1mm}
\end{align}
where we also take the circularity of $x_t[\tilde{n}]$ 
into account. By using $\tilde{N}=LN$, we can find that 
$E[{X}_{t,i}(\vartheta){X}_{t,b}^*(\vartheta)]=\sigma^2\sum_{\tilde{n}=0}^{\tilde{N}-1}e^{j2\pi\tilde{n}(\frac{b-i}{N})}=\tilde{N}\sigma^2\delta[b-i]$, 
as it is clear from~\eqref{eq:CovRycheckbarelementGaussian_noise} that $b,i\in \{0,1,\dots,N-1\}$. Hence, we can simplify~\eqref{eq:CovRycheckbarelementGaussian_noise} as
\vspace{-1mm}
\begin{align*}
&\text{Cov}[[\hat{\bf R}_{{\bar{y}}}(\vartheta)]_{m+1,m'+1},[\hat{\bf R}_{{\bar{y}}}(\vartheta)]_{a+1,a'+1}]=\sum_{t=1}^{\tau}\sum_{i=0}^{N-1}\sum_{i'=0}^{N-1} \nonumber \\
&\sum_{b=0}^{N-1}\sum_{b'=0}^{N-1}e^{\frac{j2\pi (n_m i-n_{m'} i'-n_ab+n_{a'}b')}{N}}\frac{L^2\sigma^4}{N^2\tau^2}\delta[b-i]\delta[i'-b']\nonumber \\
&=\frac{L^2\sigma^4}{N^2\tau}\sum_{i=0}^{N-1}e^{\frac{j2\pi i(n_m -n_{a})}{N}}\sum_{i'=0}^{N-1}e^{\frac{j2\pi i'(n_{a'}-n_{m'})}{N}}\nonumber\\
&=\frac{L^2\sigma^4}{\tau}\delta[m -{a}]\delta[{m'}-{a'}],\:\:\vartheta \in [0,1/N),
\vspace{-1mm}
\end{align*}
where the last equality is due to 
$n_m \in \{0,1,\dots,N-1\}$, 
for all $m$, 
and the fact that $n_m =n_{a}$ implies $m ={a}$. $\square$ 

{\color{blue}\section{Proof of~\eqref{eq:gamma_kappa}}
\label{ProofOfGammakappa}
First, by recalling that ${\bf R}_c=({\bf C}\otimes{\bf C}){\bf T}$, we can write
\begin{align}
&\gamma_\kappa =[{\bf R}_c^T{\bf R}_c]_{\kappa,\kappa}=[{\bf T}^T(({\bf C}^T{\bf C})\otimes({\bf C}^T{\bf C})){\bf T}]_{\kappa,\kappa}\nonumber \\
&=[{\bf T}^T(\text{diag}({\bf w})\otimes\text{diag}({\bf w})){\bf T}]_{\kappa,\kappa}\nonumber\\
&=\sum_{n=0}^{N-1}\sum_{n'=0}^{N-1}[{\bf T}^T]_{\kappa,Nn+n'+1}\times\nonumber\\
&[\text{diag}({\bf w})\otimes\text{diag}({\bf w})]_{Nn+n'+1,Nn+n'+1}[{\bf T}]_{Nn+n'+1,\kappa}.
\label{eq:gammakappa1}
\end{align}
Let us then recall that the $(q+1)$-th row of ${\bf T}$ is given by the $\left(\left(q-\left\lfloor\frac{q}{N}\right\rfloor\right)\text{ mod }N+1\right)$-th row of ${\bf I}_N$. We can then find that the $(\iota+1)$-th row of ${\bf T}^T$ contains ones at the $\{Nn+(n+\iota)\text{ mod }N+1\}_{n=0}^{N-1}$-th entries and zeros elsewhere. 
We can thus rewrite~\eqref{eq:gammakappa1}
\begin{align}
&\gamma_\kappa=\sum_{n=0}^{N-1}[{\bf w}\otimes{\bf w}]_{Nn+((n+\kappa-1)\text{ mod }N)+1}\nonumber\\
&=\sum_{n=0}^{N-1}[{\bf w}{\bf w}^T]_{((n+\kappa-1)\text{ mod }N)+1,n+1}\nonumber\\
&=\sum_{n=0}^{N-1}w[(n+\kappa-1)\text{ mod }N]w[n],
\end{align}
where we use ${\bf w}{\bf w}^T=\text{vec}^{-1}({\bf w}\otimes{\bf w})$ in the second equation with vec$^{-1}(.)$ the inverse of the vec$(.)$ operation.} 

\section{Proof of Theorem~3}
\label{ProofTheorem3}
To simplify the discussion, we introduce the $N^2 \times 1$ vector 
\vspace{-1mm}
\begin{equation}
\hat{\boldsymbol \rho}_{{\bar{x}}}(\vartheta)=({\bf C}\otimes{\bf C})^T\text{vec}(\hat{\bf R}_{\bar{y}}(\vartheta)).
\label{eq:rhobarx}
\vspace{-1mm}
\end{equation}
From the definition of ${\bf C}$ in Section~\ref{uncorr_bins_compression}, it is clear that the $(Nf+g+1)$-th row of $({\bf C}\otimes{\bf C})^T$ contains a single one at a certain entry and zeros elsewhere only if $f,g \in \mathcal{M}$, otherwise it contains zeros at all entries. Hence, we can write 
\begin{equation}
[\hat{\boldsymbol \rho}_{\bar{x}}(\vartheta)]_{Nf+g+1}=0, \:\:\: \text{if $ f \notin \mathcal{M}$ or  $ g \notin \mathcal{M}$}.
\label{eq:rhobarxentry}
\end{equation}
When $f,g \in \mathcal{M}$, the $(Nf+g+1)$-th entry of $\hat{\boldsymbol \rho}_{\bar{x}}(\vartheta)$ is given by one of the entries of $\text{vec}(\hat{\bf R}_{\bar{y}}(\vartheta))$.
Recall from Appendix~\ref{ProofOfGammakappa} that the $(\iota+1)$-th row of ${\bf T}^T$ contains ones at the $\{Nn+(n+\iota)\text{ mod }N+1\}_{n=0}^{N-1}$-th entries and zeros elsewhere. As a result, we can use~\eqref{rxhatbarLS},~\eqref{eq:rhobarx}, and Remark~3 
to write the $(\iota+1)$-th entry of $\hat{\bf r}_{{\bar{x}},{LS}}(\vartheta)$ in~\eqref{rxhatbarLS} as
\vspace{-1mm}
\begin{align}
&[\hat{\bf r}_{{\bar{x}},{LS}}(\vartheta)]_{\iota+1}=\frac{1}{\gamma_{\iota+1}}[{\bf T}^T\hat{\boldsymbol \rho}_{{\bar{x}}}(\vartheta)]_{\iota+1}\nonumber\\
&\:\:=\frac{1}{\gamma_{\iota+1}}\sum_{n=0}^{N-1}[\text{vec}^{-1}(\hat{\boldsymbol \rho}_{\bar{x}}(\vartheta))]_{(n+\iota)\text{ mod }N+1,n+1},
\label{eq:rbarxLSasrho}
\vspace{-1mm}
\end{align}
with $\iota=0,1,\dots,N-1$ 
and $\text{vec}^{-1}(\hat{\boldsymbol \rho}_{{\bar{x}}}(\vartheta))$ an $N\times N$ matrix. 

At this stage, let us introduce the following definition. 
\vspace{0.5mm}
\newline
\hspace*{1mm}{\it Definition 3: 
Define the collection of $[\text{vec}^{-1}(\hat{\boldsymbol \rho}_{{\bar{x}}}(\vartheta))]_{g'+1,f'+1}$ for $f',g'\in\{0,1,\dots,N-1\}$ and all $((g'-f')\text{ mod }N+1)=\kappa$ as the $\kappa$-th modular diagonal of $\text{vec}^{-1}(\hat{\boldsymbol \rho}_{{\bar{x}}}(\vartheta))$. Note that the first modular diagonal of $\text{vec}^{-1}(\hat{\boldsymbol \rho}_{{\bar{x}}}(\vartheta))$ is its main diagonal.}
\vspace{0.5mm}
\newline
We use Definition~3 to formulate the following lemma.
\vspace{0.5mm}
\newline
\hspace*{1mm}{\it Lemma~1: The $\kappa$-th modular diagonal of $\text{vec}^{-1}(\hat{\boldsymbol \rho}_{{\bar{x}}}(\vartheta))$ in~\eqref{eq:rbarxLSasrho}
contain only $\gamma_\kappa$ entries of $\text{vec}(\hat{\bf R}_{\bar{y}}(\vartheta))$ in~\eqref{eq:rhobarx}. The remaining 
$N-\gamma_\kappa$ entries of the $\kappa$-th modular diagonal of $\text{vec}^{-1}(\hat{\boldsymbol \rho}_{{\bar{x}}}(\vartheta))$ are equal to zeros.
The summation in~\eqref{eq:rbarxLSasrho} then involves $N-\gamma_{\iota+1}$ zeros and only $\gamma_{\iota+1}$ out of $M^2$ entries of $\text{vec}(\hat{\bf R}_{\bar{y}}(\vartheta))$.} 
\newline\hspace*{1mm}{\it Proof:}
Recall that, when $f,g \in \mathcal{M}$, the $(Nf+g+1)$-th entry of $\hat{\boldsymbol \rho}_{\bar{x}}(\vartheta)$ in~\eqref{eq:rbarxLSasrho} is given by one of the entries of $\text{vec}(\hat{\bf R}_{\bar{y}}(\vartheta))$. Since Remark~3 indicates that the number of pairs $g,f \in \mathcal{M}$ that lead to $(g-f)\text{ mod }N+1=\kappa$ is equal to $\gamma_{\kappa}$, it is clear from Definition~3 that the $\kappa$-th modular diagonal of $\text{vec}^{-1}(\hat{\boldsymbol \rho}_{\bar{x}}(\vartheta))$ only contains $\gamma_{\kappa}$ entries of $\text{vec}(\hat{\bf R}_{\bar{y}}(\vartheta))$. Equation~\eqref{eq:rhobarxentry} then confirms that the remaining $N-\gamma_{\kappa}$ entries of the $\kappa$-th modular diagonal of $\text{vec}^{-1}(\hat{\boldsymbol \rho}_{\bar{x}}(\vartheta))$ 
are equal to zero. 
Next, observe that the summation in~\eqref{eq:rbarxLSasrho} is the sum of all terms in the $(\iota+1)$-th modular diagonal of $\text{vec}^{-1}(\hat{\boldsymbol \rho}_{\bar{x}}(\vartheta))$. This can be found by applying Definition~3 on the column and row indices of $\text{vec}^{-1}(\hat{\boldsymbol \rho}_{\bar{x}}(\vartheta))$ in~\eqref{eq:rbarxLSasrho}, i.e.,
\vspace{-1.5mm}
\begin{eqnarray}
&((n+\iota)\text{ mod }N-n)\text{ mod }N+1\nonumber\\
&=(n+\iota-n)\text{ mod }N+1=\iota+1,\nonumber
\vspace{-1.5mm}
\end{eqnarray}
which exploits the property that 
$(\kappa \text{ mod }N+\kappa')\text{ mod }N=(\kappa+\kappa')\text{ mod }N$. 
This concludes the proof. $\square$

Let us now define ${\boldsymbol \Sigma}_{\hat{\rho}_{{\bar{x}}}}(\vartheta)$ as the $N^2\times N^2$ covariance matrix of $\hat{\boldsymbol \rho}_{{\bar{x}}}(\vartheta)$ in~\eqref{eq:rhobarx}, which can be written as ${\boldsymbol \Sigma}_{\hat{\rho}_{{\bar{x}}}}(\vartheta)=({\bf C}\otimes{\bf C})^T
{\boldsymbol \Sigma}_{\hat{R}_{\bar{y}}}(\vartheta)({\bf C}\otimes{\bf C})$. First, recall~\eqref{eq:rhobarxentry} and that when $f,g \in \mathcal{M}$, the $(Nf+g+1)$-th entry of $\hat{\boldsymbol \rho}_{\bar{x}}(\vartheta)$ in~\eqref{eq:rbarxLSasrho} is given by one of the entries of $\text{vec}(\hat{\bf R}_{\bar{y}}(\vartheta))$. By also recalling that, for circular complex Gaussian i.i.d. noise $x_t[\tilde{n}]$, ${\boldsymbol \Sigma}_{\hat{R}_{\bar{y}}}(\vartheta)$ is a diagonal matrix whose elements are given by~\eqref{eq:CovRycheckbarelementGaussian_noise_propos}, we can find that ${\boldsymbol \Sigma}_{\hat{\rho}_{{\bar{x}}}}(\vartheta)$ is also a diagonal matrix with its diagonal elements given by
\begin{equation}
[\text{diag}({\boldsymbol \Sigma}_{\hat{\rho}_{{\bar{x}}}}(\vartheta))]_{Nf+g+1}=
\left\{ \begin{array}{ll} 
\frac{L^2\sigma^4}{\tau},\quad\text{if $f,g \in \mathcal{M}$.}  \\
0, \: \text{if $ f \notin \mathcal{M}$ or  $ g \notin \mathcal{M}$.}
\end{array} \right.
\label{eq:diag_Cov_rho}
\end{equation}
By taking~\eqref{eq:rbarxLSasrho},~\eqref{eq:diag_Cov_rho}, and the diagonal structure of ${\boldsymbol \Sigma}_{\hat{\rho}_{\bar{x}}}(\vartheta)$ into account, we can then write the entry of ${\boldsymbol \Sigma}_{\hat{r}_{{\bar{x}},LS}}(\vartheta)$ in~\eqref{eq:Covar_scheckhatbarXLS} 
at the $(\iota+1)$-th row and the $(\iota'+1)$-th column as 
\vspace{-1mm}
\begin{align}
&\text{Cov}[[\hat{\bf r}_{{\bar{x}},{LS}}(\vartheta)]_{\iota+1},[\hat{\bf r}_{{\bar{x}},{LS}}(\vartheta)]_{\iota'+1}]
=\frac{1}{\gamma_{\iota+1}\gamma_{\iota'+1}}\times\nonumber\\
&\sum_{n=0}^{N-1}\sum_{n'=0}^{N-1}\left\{[{\bf T}^T]_{\iota+1,Nn+n'+1}[{\boldsymbol \Sigma}_{\hat{\rho}_{{\bar{x}}}}(\vartheta)]_{Nn+n'+1,Nn+n'+1}\times\right.\nonumber\\
&\quad\quad\quad\quad\left.[{\bf T}]_{Nn+n'+1,\iota'+1}\right\}
=\frac{\delta[\iota-\iota']}{\gamma^2_{\iota+1}}\times\nonumber\\
&\sum_{n=0}^{N-1}[{\boldsymbol \Sigma}_{\hat{\rho}_{{\bar{x}}}}(\vartheta)]_{Nn+
((n+\iota)\text{ mod }N)+1,Nn+
((n+\iota)\text{ mod }N)+1},
\label{eq:Entryofscheckbar_x_LS_gaussiannoise}	
\vspace{-1mm}
\end{align}
for $\iota,\iota'=0,1,\dots,N-1$, which implies that ${\boldsymbol \Sigma}_{\hat{r}_{{\bar{x}},LS}}(\vartheta)$ is also a diagonal matrix for circular complex Gaussian i.i.d. noise $x_t[\tilde{n}]$. 
Recall from the proof of Lemma~1 
that the summation in 
\eqref{eq:rbarxLSasrho} is the sum of all terms in the $(\iota+1)$-th modular diagonal of $\text{vec}^{-1}(\hat{\boldsymbol \rho}_{{\bar{x}}}(\vartheta))$. We can then observe that the summation in~\eqref{eq:Entryofscheckbar_x_LS_gaussiannoise} is the sum of the variance of each term in the $(\iota+1)$-th modular diagonal of $\text{vec}^{-1}(\hat{\boldsymbol \rho}_{{\bar{x}}}(\vartheta))$. 
Using Lemma~1 
and~\eqref{eq:diag_Cov_rho}, we can rewrite~\eqref{eq:Entryofscheckbar_x_LS_gaussiannoise} as 
\begin{equation}
\text{Cov}[[\hat{\bf r}_{{\bar{x}},{LS}}(\vartheta)]_{\iota+1},[\hat{\bf r}_{{\bar{x}},{LS}}(\vartheta)]_{\iota'+1}]
=\frac{L^2\sigma^4}{\gamma_{\iota+1}\tau}\delta[\iota-\iota'],
\label{eq:Entryofscheckbar_x_LS_gaussiannoise2}
\end{equation}
for $\iota,\iota'=0,1,\dots,N-1$. By considering~\eqref{eq:Covar_ScheckhatXLS} and noticing that $[{\bf B}^T\otimes{\bf B}^{H}]_{Ni+i'+1,Nn+n'+1}$ $=\frac{1}{N^2}e^{-j\frac{2\pi}{N}(n'i'-ni)}$, let us rewrite $\text{Var}[\hat{P}_{{x},{LS}}(\vartheta+\frac{i}{N})]$ in~\eqref{eq:Var_LS_periodogoram}, for $\vartheta \in [0,1/N)$ and $i=0,1,\dots,N-1$, as
\begin{align}
&\text{Var}[\hat{P}_{{x},{LS}}(\vartheta +\frac{i}{N})]
=\frac{N^4}{\tilde{N}^2}\sum_{n=0}^{N-1}\sum_{n'=0}^{N-1}\sum_{\nu=0}^{N-1}\sum_{\nu'=0}^{N-1}\nonumber \\
&\left\{[{\bf B}^T\otimes{\bf B}^{H}]_{Ni+i+1,Nn+n'+1}\times\right.\nonumber\\
&\left.[{\bf T}{\boldsymbol \Sigma}_{\hat{r}_{{\bar{x}},LS}}(\vartheta){\bf T}^T]_{Nn+n'+1,N\nu+\nu'+1}[{\bf B}^*\otimes{\bf B}]_{N\nu+\nu'+1,Ni+i+1}\right\}\nonumber \\
&=\frac{1}{L^2N^2}\sum_{n=0}^{N-1}\sum_{n'=0}^{N-1}\sum_{\nu=0}^{N-1}\sum_{\nu'=0}^{N-1}\left\{
e^{-j\frac{2\pi}{N}i(n'-n+\nu-\nu')}\times\right.\nonumber\\
&\left.[{\bf T}{\boldsymbol \Sigma}_{\hat{r}_{{\bar{x}},LS}}(\vartheta){\bf T}^T]_{Nn+n'+1,N\nu+\nu'+1}\right\}.
\label{eq:VarPxLSwhitenoise2}
\end{align}
We now recall that the $(q+1)$-th row of ${\bf T}$ is given by the $\left(\left(q-\left\lfloor\frac{q}{N}\right\rfloor\right)\text{ mod }N+1\right)$-th row of ${\bf I}_N$, exploit the diagonal structure of ${\boldsymbol \Sigma}_{\hat{r}_{{\bar{x}},LS}}(\vartheta)$ for circular complex Gaussian i.i.d. noise $x_t[\tilde{n}]$, and use~\eqref{eq:Entryofscheckbar_x_LS_gaussiannoise2} to write 
\begin{align}
&[{\bf T}{\boldsymbol \Sigma}_{\hat{r}_{{\bar{x}},LS}}(\vartheta){\bf T}^T]_{Nn+n'+1,N\nu+\nu'+1}\nonumber\\
&=\frac{L^2\sigma^4}{\tau}\sum_{\iota=0}^{N-1}\frac{1}{\gamma_{\iota+1}}[{\bf T}]_{Nn+n'+1,\iota+1}[{\bf T}^T]_{\iota+1,N\nu+\nu'+1} \nonumber \\
&=\frac{L^2\sigma^4}{\tau}\frac{\delta[(n'-n)\text{ mod }N-(\nu'-\nu)\text{ mod }N]}{\gamma_{(n'-n)\text{ mod }N+1}},
\label{eq:TSigma_sT}
\end{align}
for $n,n',\nu,\nu'=0,1,\dots,N-1$. By inserting~\eqref{eq:TSigma_sT} into~\eqref{eq:VarPxLSwhitenoise2}, 
the variance of 
$\hat{P}_{{x},{LS}}(\vartheta+\frac{i}{N})$, for circular complex Gaussian i.i.d. noise $x_t[\tilde{n}]$ 
and $i=0,1,\dots,N-1$, is given by
\begin{align*}
&\text{Var}[\hat{P}_{{x},{LS}}(\vartheta+\frac{i}{N})]
=\frac{1}{L^2N^2}\sum_{n=0}^{N-1}\sum_{n'=0}^{N-1}\frac{L^2\sigma^4N}{\tau\gamma_{(n'-n)\text{ mod }N+1}}\nonumber\\
&=\frac{\sigma^4}{\tau}\sum_{n=0}^{N-1}\frac{1}{\gamma_{n+1}}
=\frac{\sigma^4}{M\tau}+\frac{\sigma^4}{\tau}\sum_{n=1}^{N-1}\frac{1}{\gamma_{n+1}},\:\vartheta \in [0,1/N),
\end{align*}
where we use the last part of Remark~3 in the last equality. $\square$

\end{document}